\documentclass[a4paper,notitlepage,twocolumn,footnoteinbib,superscriptaddress,10pt]{revtex4-1}
\usepackage{amsmath,bbold,amssymb,amsthm,mathtools,color,listings,graphicx}
\usepackage[colorlinks=true,linkcolor=blue,citecolor=blue]{hyperref}

\newcommand{\revone}[1]{\textcolor{black}{#1}}
\newcommand{\revtwo}[1]{\textcolor{black}{#1}}
\newcommand{\revthree}[1]{\textcolor{black}{#1}}

\begin{document}

\title{Nonlocality-induced surface localization in Bose-Einstein condensates of light}
\author{Marcello Calvanese Strinati}
\email{marcello.calvanesestrinati@gmail.com}
\affiliation{Centro Ricerche Enrico Fermi (CREF), Via Panisperna 89a, 00184 Rome, Italy}
\affiliation{Physics Department, Sapienza University of Rome, 00185 Rome, Italy}
\author{Frank Vewinger}
\affiliation{Institut f\"ur Angewandte Physik, Universit\"at Bonn, Wegelerstra{\ss}e 8, 53115 Bonn, Germany}
\author{Claudio Conti}
\affiliation{Physics Department, Sapienza University of Rome, 00185 Rome, Italy}
\affiliation{Institute for Complex Systems, National Research Council (ISC-CNR), 00185 Rome, Italy}
\affiliation{Centro Ricerche Enrico Fermi (CREF), Via Panisperna 89a, 00184 Rome, Italy}
\date{\today}

\begin{abstract}
The ability to create and manipulate strongly correlated quantum many-body states is of central importance to the study of collective phenomena in several condensed-matter systems. In the last decades, a great amount of work has been focused on ultracold atoms in optical lattices, which provide a flexible platform to simulate peculiar phases of matter both for fermionic and bosonic particles. The recent experimental demonstration of Bose-Einstein condensation (BEC) of light in dye-filled microcavities has opened the intriguing possibility to build photonic simulators of solid-state systems, with potential advantages over their atomic counterpart. A distinctive feature of photon BEC is the thermo-optical nature of the effective photon-photon interaction, which is intrinsically nonlocal and can thus induce interactions of arbitrary range. This offers the opportunity to systematically study the collective behaviour of many-body systems with tunable interaction range. In this paper, we theoretically study the effect of nonlocal interactions in photon BEC. We first present numerical results of BEC in a double-well potential, and then extend our analysis to a short one-dimensional lattice with open boundaries. By resorting to a numerical procedure inspired by the Newton-Raphson method, we simulate the time-independent Gross-Pitaevskii equation and provide evidence of surface localization induced by nonlocality, where the condensate density is localized at the boundaries of the potential. Our work paves the way towards the realization of synthetic matter with photons, where the interplay between long-range interactions and low dimensionality can lead to the emergence of unexplored nontrivial collective phenomena.
\end{abstract}

\maketitle

\section{Introduction}
\label{sec:introduction}
The study of exotic phenomena arising from the interplay between low dimensionality and strong interactions has been a very active field of research in the last decades, ranging from electronic and magnetic properties in low dimension~\cite{baeriswyl2007strong} to the emergence of topologically ordered states~\cite{RevModPhys.89.041004}. While the discovery and characterization of such phenomena was first made in solid-state systems, the recent advances in the fields of ultracold atoms~\cite{RevModPhys.91.015005}, \revtwo{ultracold dipolar molecules~\cite{hazzard2018}}, and photonics~\cite{RevModPhys.85.299,RevModPhys.91.015006} have granted the possibility to engineer alternative physical systems to simulate solid-state matter. The development of such an artificial matter, often referred to as \emph{synthetic matter}, offers the ability to manipulate with unprecedented precision the physics of the simulated material, e.g., interactions and presence of artificial magnetic-orbital fields~\cite{RevModPhys.83.1523,RevModPhys.91.015006}, allowing us to unveil properties and phases of matter in physical conditions that are difficult to access in solid-state materials~\cite{ozawa2019topologicalsyntheticmatter}.

In this framework, the recent experimental demonstration of two-dimensional Bose-Einstein condensation (BEC) of light in dye-filled microcavities has opened the possibility to use photons to simulate the equilibrium properties of bosonic matter. First observed in 2010~\cite{weitz2010becphotons1,weitz2010becphotons2}, photon BEC is an out-of-equilibrium state of light showing effective thermal steady-state properties, achieved by pumping photons into a dye-filled microcavity in proper conditions. The geometry of the cavity mirrors provides an effective trapping potential and mass to photons, while the pumping sets their chemical potential. Specifically, the effective trapping potential is induced by a proper shaping of the cavity mirrors, which were curved in Refs.~\cite{weitz2010becphotons1,weitz2010becphotons2} to provide an harmonic confinement (a later experiment on the same system was reported also in Ref.~\cite{PhysRevA.91.033813}). The effective mass results from the freezing of the longitudinal quantum number of the cavity modes populated by the photons reemitted by the dye molecules. This is obtained by using a short distance between the cavity mirrors to have a free spectral range that is comparable with the spectral width of the emission line of the dye solution within the cavity. Thermalization of the photon gas and the possibility of a non-vanishing chemical potential is achieved by repeated absorption and emission of photons by the dye molecules pumped with an external laser, with consequent conservation of the average number of photons. In these conditions, albeit the system is out of equilibrium, the resulting state of the emitted light can be seen as effectively in a steady state close to equilibrium~\cite{PhysRevLett.108.160403,PhysRevLett.113.135301}, and described by a wavefunction $\Psi(\mathbf{r})$ and chemical potential $\mu$, quantifying respectively the electric field and the energy stored in the condensate optical mode, obeying a proper Gross-Pitaevskii equation~\cite{weitz2010becphotons1,PhysRevA.89.033844,PhysRevA.90.043853,weitz2017thermoopticalinteraction,Stein_2022}. 

Recently, in 2019, a photon BEC in a double-well potential was experimentally obtained~\cite{doi:10.1126/science.aay1334} using similar methods to those in Refs.~\cite{weitz2010becphotons1,weitz2010becphotons2}. The effective two-well potential for photons was imprinted by properly shaping the cavity mirrors using dedicated delamination techniques. Importantly, these techniques can be used to imprint any potential, e.g., lattices or non-periodic potentials~\cite{weitzarbitrarypotential2017,weitzarbitrarypotential2020,Walker:21}. The experimental demonstration of photon BEC in a double well, together with the ability to realize arbitrary lattice potentials, has opened the possibility to use photons BEC to realize synthetic matter at room temperature, envisioning the study of many-body phenomena in parameter regimes that are often unexplored in cold-atom experiments. In this respect, a peculiarity of photon BEC is the nature of the effective photon-photon interaction, which arises from a thermo-optical nonlinearity and it is thus intrinsically nonlocal~\cite{PhysRevA.90.043853,weitz2017thermoopticalinteraction,Stein_2019}: The condensate intensity heats up the medium at some point $\mathbf{r}$, causing a spatial modification of the index of refraction that depends nonlocally on the intensity via a proper Green's function. As such, the interaction strength at a given point $\mathbf{r}$ depends on the wavefunction of the whole condensate, thereby manifesting its nonlocal (i.e., long-range) nature. \revthree{In addition to the nonlocal thermo-optical nonlinearity, also local attractive interactions based on the optical Kerr effect are present. In current experiments, these interactions are much weaker than the thermo-optical one, and their influence on the photon BEC is thus negligible}. As of today, experiments and theoretical analyses of photon BECs are performed considering effective photon-photon interactions with very short range. In the perspective of using arrays of photon BECs as simulators of complex many-body quantum states, a deeper knowledge on how the presence \revone{of} strong long-range interactions affects the condensate properties is highly desirable.

In this paper, we analytically and \revone{numerically} study the effect of strong interactions with arbitrary range in photon BEC. We first present our results on the two-well potential of Ref.~\cite{doi:10.1126/science.aay1334}, and then extend our analysis to a short lattice of six sites with open boundaries. We numerically determine the condensate wavefunction and corresponding chemical potential by solving the time-independent Gross-Pitaevskii equation with nonlocal nonlinearity~\cite{PhysRevA.90.043853}, resorting to a numerical algorithm inspired by the Newton-Raphson method~\cite{ortega1970iterative,press2002numerical}. We simulate the photon BEC in different regimes of interaction, and find that the nonlocal nonlinearity can induce a level inversion in the low-energy spectrum, remarkably stabilizing a ground state where the condensate density localizes at the boundaries of the potential.

This paper is organized as follows. In Sec.~\ref{sec:doublewell}, we study the effect of the nonlocal \revone{nonlinearity} in the double-well potential. In Sec.~\ref{sec:lattices}, we extend our discussion to a short one-dimensional lattice, providing evidence of the onset of surface localization of the condensate density, in proper nonlocal regimes. We draw our conclusions in Sec.~\ref{sec:conclusions}, and provide additional details in the appendixes.

\section{BEC in a double-well potential}
\label{sec:doublewell}
We open by discussing the effect of a nonlocal interaction in the case of the BEC of photons in the double-well potential~\cite{doi:10.1126/science.aay1334}. We first introduce our model and notations, and then present our numerical results.

\subsection{Model}
\revtwo{In photon BEC,} the effective steady state \revtwo{is described} by a \revtwo{time-independent} nonlocal Gross-Pitaevskii equation in two dimensions~\cite{weitz2010becphotons1,PhysRevA.90.043853,weitz2017thermoopticalinteraction}
\begin{equation}
\left[-\frac{\hbar^2}{2m}\nabla^2\!+\!V(\mathbf{r})\!+\!\int\! d\mathbf{r}'\,G(\mathbf{r},\mathbf{r}'){\left|\Psi\left(\mathbf{r'}\right)\right|}^2\right]\!=\!\mu\Psi(\mathbf{r}) \,\, ,
\label{eq:grosspitaevskiiequation1}
\end{equation}
where $\mathbf{r}=(x,y)$ denotes the two-dimensional spatial coordinate, $\nabla^2=\partial^2/\partial x^2+\partial^2/\partial y^2$, and $V(\mathbf{r})$ is the effective potential. The wavefunction $\Psi(\mathbf{r})$ is normalized as $\int d\mathbf{r}\,|\Psi(\mathbf{r})|^2=N_0$, where $N_0$ is the total (average) number of photons in the condensate mode, and the chemical potential $\mu$ encodes the energy stored in this mode. The Green's function \revone{$G(\mathbf{r},\mathbf{r}')$} describes the effective nonlocal photon-photon thermo-optical interaction~\cite{PhysRevA.90.043853}. \revone{In the following, we consider an isotropic interaction, i.e., $G(\mathbf{r},\mathbf{r}')\equiv G(\mathbf{r}-\mathbf{r}')$}. We can equivalently recast Eq.~\eqref{eq:grosspitaevskiiequation1} by rescaling the wavefunction as $\Psi(\mathbf{r})=\sqrt{N_0}\psi(\mathbf{r})$, where now $\psi(\mathbf{r})$ has unit norm, and by introducing the \revone{normalized} interaction kernel $K(\mathbf{r})=G(\mathbf{r})/g$ \revone{where} $g\coloneqq\int d\mathbf{r}\,G(\mathbf{r})$ \revone{quantifies the interaction strength}, i.e.,
\begin{equation}
\left[-\frac{\hbar^2}{2m}\nabla^2\!+\!V(\mathbf{r})\!+gN_0\!\!\int\! d\mathbf{r}'\,K(\mathbf{r},\mathbf{r}'){\left|\psi\left(\mathbf{r'}\right)\right|}^2\right]\!=\!\mu\psi(\mathbf{r}) \,\, .
\label{eq:grosspitaevskiiequation2}
\end{equation}
The sign of $g$ in Eq.~\eqref{eq:grosspitaevskiiequation2} determines the nature of the interaction. It is repulsive, or defocusing, for $g>0$, while it is attractive, or focusing, for $g<0$. From now on, we focus on the case of repulsive interaction $g>0$.

The goal now is to study numerically the effect of the nonlocal interaction on the condensate wavefunction. To reach this goal, we restrict ourselves to one dimension ($\mathbf{r}=x$ and $\nabla^2=d^2/dx^2$). This choice allows us to drastically reduce the numerical complexity of the problem while granting access to the most relevant informations on the interplay between the interaction strength and its range, encoded in $K(x,x')$ as detailed below. In our simulations, we model the double-well potential as
\begin{equation}
V_{\rm 2W}(x)\!=\!\left\{
\begin{array}{ll}
\!\!\cfrac{m\omega^2}{2}\displaystyle{{\left(\cfrac{x_{\rm min}}{\pi}\right)}^2\!\!\cos^2\!\left(\cfrac{\pi x}{x_{\rm min}}\right)} &  (|x|<x_{\rm min}/2)\\\\
\!\!\displaystyle{\cfrac{m\omega^2}{2}{\left(|x|-\frac{x_{\rm min}}{2}\right)}^2} & (|x|\geq x_{\rm min}/2)
\end{array}
\right. \,\, ,
\label{eq:equationdoublewellpotential}
\end{equation}
where the subscript ``2W'' stands for ``double-well''. In Eq.~\eqref{eq:equationdoublewellpotential}, $\omega$ is the effective trapping potential on each minimum, and $x_{\rm min}$ is the distance between the two minima. In the case of local interaction, the kernel is a delta function $K(x)=\delta(x)$. \revone{For} a nonlocal interaction, \revone{$K(x)$ describes heat transport, and its form is found by solving the appropriate heat diffusion equation~\cite{PhysRevA.90.043853,weitz2017thermoopticalinteraction,Stein_2022}}. \revtwo{Here, to limit the numerical complexity, we} model the kernel as a regularized box potential \revtwo{of the form}
\begin{equation}
K(x)=K_0\left[\tanh\left(\frac{x+\sigma}{w}\right)-\tanh\left(\frac{x-\sigma}{w}\right)\right] \,\, ,
\label{eq:equationinteractionkernel}
\end{equation}
where $\sigma$ denotes the \revtwo{effective thermo-optic} interaction range (i.e., $2\sigma$ is approximately the size of the box), $w$ quantifies how sharply the interaction goes to zero around the box boundaries $x=\pm\sigma$, and $K_0=1/\int_{-\infty}^{\infty} dx\,K(x)$ ensures the kernel unit normalization (for sufficiently small $w$, $K_0\simeq1/2\sigma$). \revone{The choice of the box potential in Eq.~\eqref{eq:equationinteractionkernel} allows us to systematically study the effect of nonlocality while keeping a reasonable numerical complexity of our simulations (details are given in Appendix~\ref{appendix:detailsonthenumericalsimulations})}. \revone{Hereafter, we} refer to the case \revone{$K(x)=\delta(x$)} of local interaction as $\sigma=0$.

\begin{figure}[t]
\centering
\includegraphics[width=4.5cm]{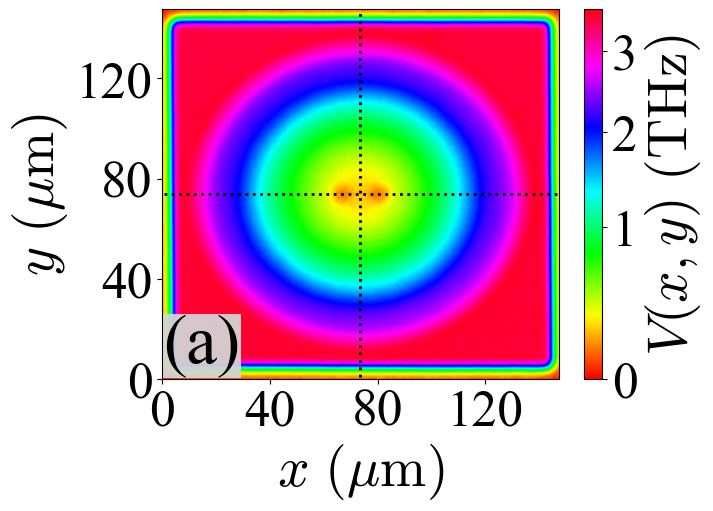}
\includegraphics[width=4cm]{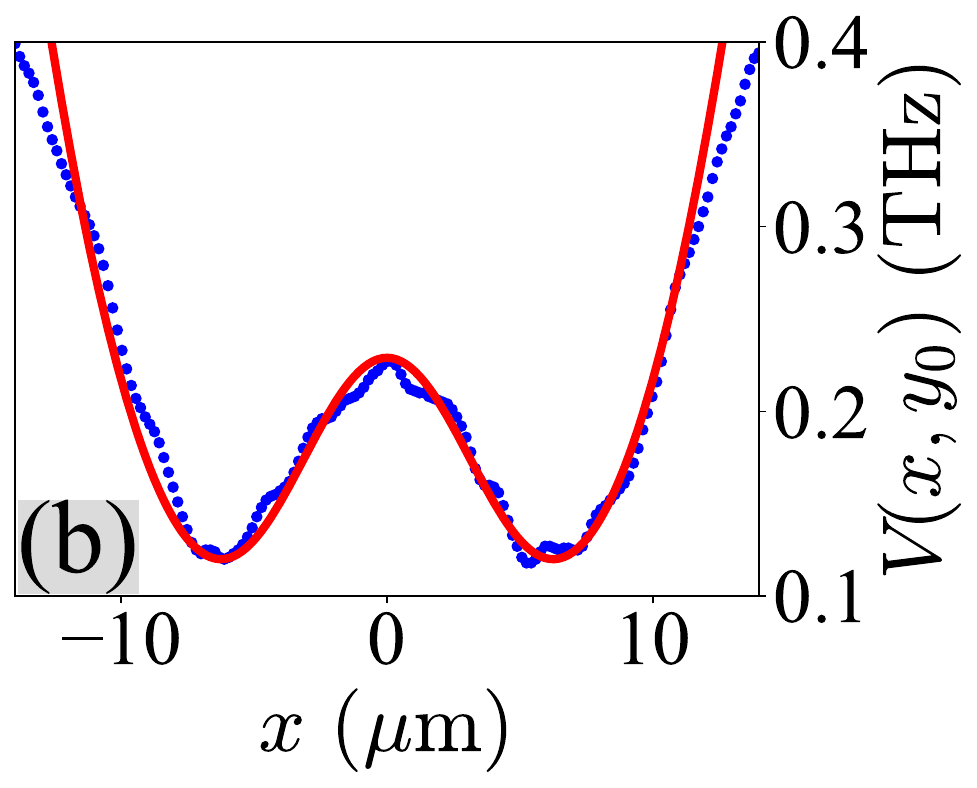}
\caption{(a) Colormap of the experimental data for the double-well potential $V(x,y)$ from Ref.~\cite{doi:10.1126/science.aay1334}. (b) Cut along the $x$-axis of the data in panel (a) (blue points) for $y_0=73.675\,{\rm\mu m}$, marked by the dotted horizontal black line, and fit with the function in Eq.~\eqref{eq:equationdoublewellpotential} using $\sqrt{m}\omega=0.117346\,{\rm \sqrt{THz}/\mu m}$ \revone{(i.e., $\omega\simeq\revone{1.08}\times10^{12}\,{\rm rad/s}$ with $m\simeq\revone{7.76}\times10^{-36}\,{\rm kg}$)} and $x_{\rm min}=12.5\,{\rm \mu m}$ (red line, see Appendix~\ref{appendix:detailsonthefitdoublewellpotential}). The origin of the $x$-axis in the panel (b) has been shifted to have a symmetric double well around $x=0$ [which corresponds to $x=73.5\,{\rm\mu m}$ in panel (a), marked by the vertical black dotted line].}
\label{fig:potentialdoublewellweitz}
\end{figure}

In Eq.~\eqref{eq:equationdoublewellpotential}, we have two independent length scales: $\xi=\sqrt{\hbar/m\omega}$, which is the characteristic size of the wavefunction density within each well, and the distance between the minima of the two wells $x_{\rm min}$. To use experimentally meaningful values in Eq.~\eqref{eq:equationdoublewellpotential}, we take the photon mass as $m\simeq\revone{7.76}\times10^{-36}\,{\rm kg}$ and fit the low-energy part of the potential from Ref.~\cite{doi:10.1126/science.aay1334}, in particular the cut along the axis of the double well, using the function in Eq.~\eqref{eq:equationdoublewellpotential}. The result of the fit is shown in Fig.~\ref{fig:potentialdoublewellweitz}. From the experimental data, we extract the distance between the two minima $x_{\rm min}\simeq12.5\,{\rm \mu m}$, and estimate the trapping frequency \revtwo{in a single microsite as} $\omega\simeq\revone{1.08}\times10^{12}\,{\rm rad/s}$, from which $\xi=\sqrt{\hbar/m\omega}\simeq\revone{3.54}\,{\rm \mu m}$ follows (see Appendix~\ref{appendix:detailsonthefitdoublewellpotential} for details). This allows to rewrite Eq.~\eqref{eq:grosspitaevskiiequation2} in dimensionless units, using $\xi$ and $\hbar\omega$ as characteristic length and energy scales, respectively.

\subsection{Numerical method}
We then simulate Eq.~\eqref{eq:grosspitaevskiiequation2} in dimensionless units by resorting to a numerical method inspired by the Newton-Raphson method~\cite{ortega1970iterative,press2002numerical}, which is detailed in Appendix~\ref{appendix:detailsonthenumericalsimulations}. We here report the main steps for the sake of clarity. Our goal is to find the condensate wavefunction $\psi(x)$, and associated chemical potential $\mu$, for given values of $gN_0$ and $\sigma$, using the kernel in Eq.~\eqref{eq:equationinteractionkernel}. We discretize the $x$-axis using a grid of $M$ points. Since the double-well potential confines the wavefunction within a finite segment of length $S$ along the $x$-axis, symmetric with respect to $x=0$, we can truncate the $x$-domain into a box of linear length $L_x>S$, where $L_x$ is chosen such that the wavefunction is zero for all $x$ outside the box. Then, we discretize $x\equiv x_j=-L_x/2+(j-1)\Delta_x$, with $\Delta_x=L_x/M$, and $j=1,\ldots,M$. Consequently, the wavefunction $\psi(x)$ is in turn discretized into a $M$-dimensional vector $\vec\psi$, whose components are $\psi_j=\psi(x_j)$. \revone{Here, $S$ is basically the spatial extension along $x$ of the condensate density, which is approximately $S\simeq x_{\rm min}$ due to the high potential barriers for $|x|>x_{\rm min}/2$.}

We identify two parts in the Hamiltonian in Eq.~\eqref{eq:grosspitaevskiiequation2}: The linear Hamiltonian $H_0=-(1/2)d^2/dx^2+V(x)$, and the nonlinear contribution $F(x)=gN_0\int dx'\,K(x-x')|\psi(x')|^2$. The linear Hamiltonian in the discrete formulation is a $M\times M$ matrix, whose exact diagonalization yields the spectrum of the Schr\"odinger's equation, i.e., the wavefunctions $\{\psi^{(0)}_m(x)\}$ and associated energies $\{\mu^{(0)}_m\}$ ($m=1,\ldots,M$) for $gN_0=0$ denoted by the superscript ``$(0)$''. The key idea behind our method is to determine, for a fixed value of $\sigma$, the nonlinear wavefunction $\psi^{(g)}_m(x)$ and corresponding chemical potential $\mu^{(g)}_m$ for a given $gN_0\neq0$ perturbatively, starting from the knowledge of $\psi^{(0)}_m(x)$ and $\mu^{(0)}_m$. Then, starting from $gN_0=0$, we gradually ramp up the value of $gN_0$ by a small step $dg$: At the \revone{$q$-th} step of this ramping, we determine $\psi^{(q\,dg)}_m(x)$ as $\psi^{(q\,dg)}_m(x)=\psi^{((q-1)dg)}_m(x)+\varphi(x)$, for a proper small correction $\varphi(x)$ that is found as detailed in Appendix~\ref{appendix:detailsonthenumericalsimulations}. This procedure is \revone{seeded using $\psi^{(0)}_m(x)$ and $\mu^{(0)}_m$ for $q=1$, and} repeated \revone{by increasing $q$} until the value of $gN_0$ appearing in Eq.~\eqref{eq:grosspitaevskiiequation2} is reached.

\begin{figure*}[t]
\centering
\includegraphics[width=5.91cm]{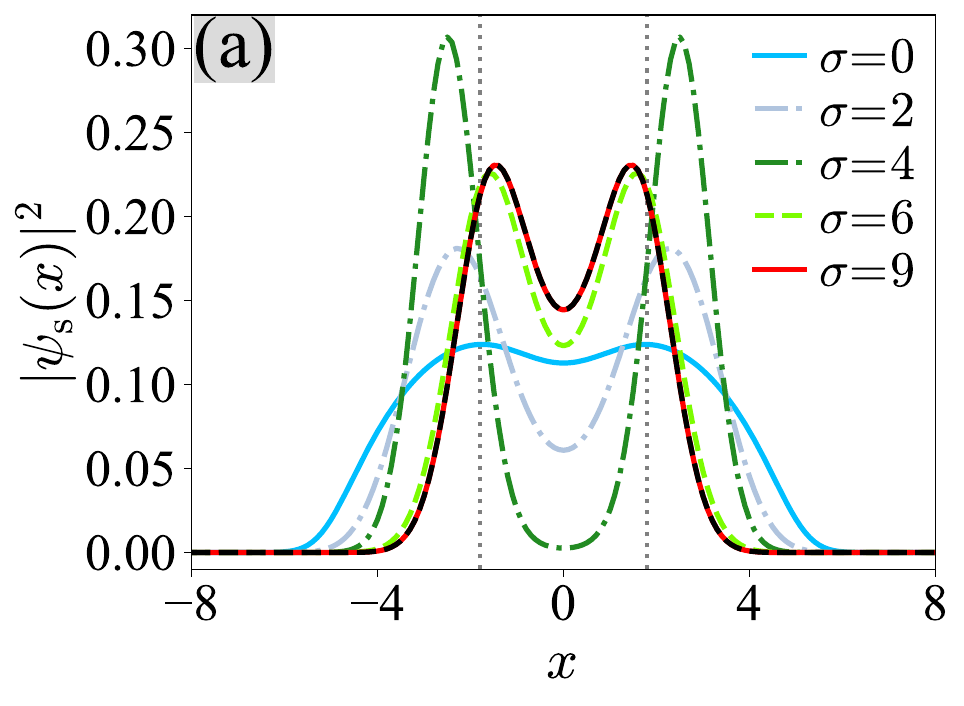}
\includegraphics[width=5.91cm]{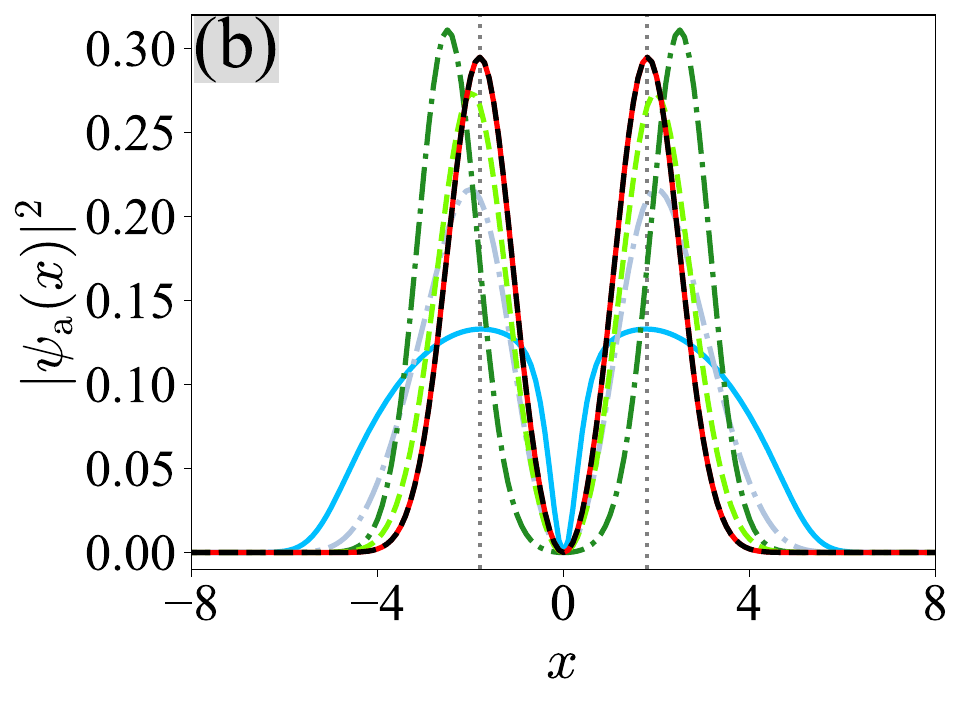}
\includegraphics[width=5.85cm]{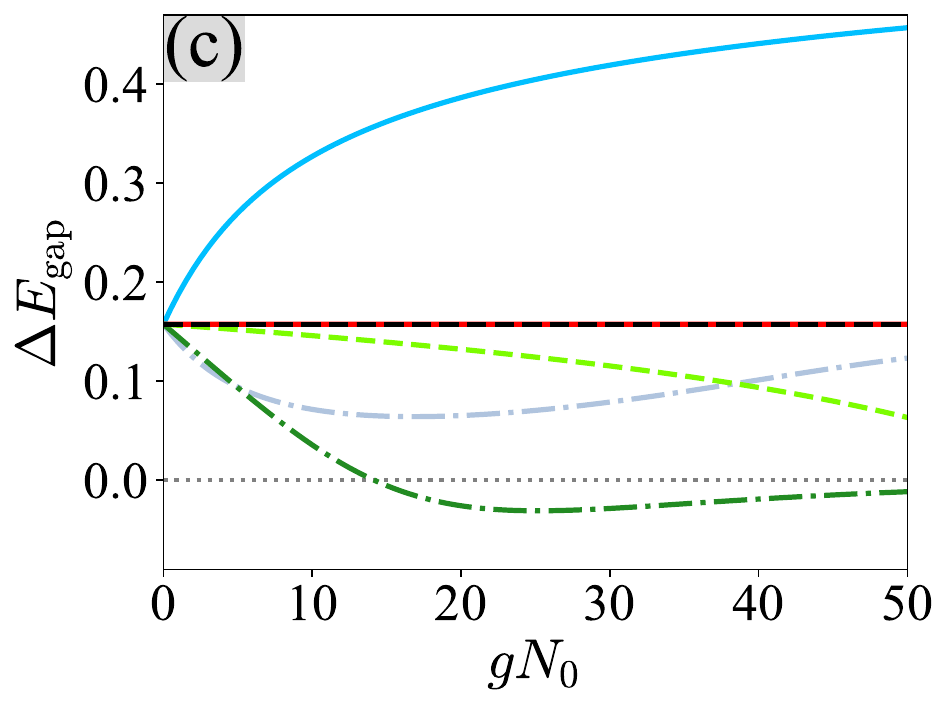}
\caption{Condensate density as a function of $x$ (in \revthree{units of $\xi$}, see text) from the simulation of the Gross-Pitaevskii equation for (a) symmetric $|\psi_{\rm s}(x)|^2$, and (b) antisymmetric state $|\psi_{\rm a}(x)|^2$, with the double-well potential in Eq.~\eqref{eq:equationdoublewellpotential} (see also Fig.~\ref{fig:potentialdoublewellweitz}b). \revthree{The condensate density at fixed $gN_0=50$ is plotted for $\sigma=0$ (blue solid line), $\sigma=2$ (blue-gray dash-dotted line), $\sigma=4$ (dark green dash-dotted line), $\sigma=6$ (light green dashed line), and $\sigma=9$ (solid red line)}. For clarity, we plot also the condensate density in the linear case (black dashed line). The numerical parameters are: $L_x=40$, $M=400$, $x_{\rm min}=3.53121$, $w=1$, and $dg=0.5$. Vertical gray dashed lines mark the position of the potential minima $x=\pm x_{\rm min}/2$. (c) Energy gap $\Delta E_{\rm gap}\coloneqq \mu^{(g)}_{\rm a}-\mu^{(g)}_{\rm s}$ \revthree{in units of $\hbar\omega$} as a function of $gN_0$. Different curves refer to different values of $\sigma$, with color coding as in panels (a) and (b). The horizontal dotted gray line marks the value $\Delta E_{\rm gap}=0$, below which the antisymmetric state becomes the lowest-energy state of the double well. This \revthree{happens for $\sigma=4$ and} sufficiently large $gN_0$, signaling the preference for the system to create a full depletion area (with a node) between the two wells.}
\label{fig:doublewellpotentialwavefunction}
\end{figure*}

\subsection{Numerical results}
Using this approach, we numerically simulate Eq.~\eqref{eq:grosspitaevskiiequation2} using $L_x=40$, $M=400$ (i.e., $\Delta_x=0.1$), and $x_{\rm min}=3.53121$. We set $w=1$ in Eq.~\eqref{eq:equationinteractionkernel} and ramp up the coupling strength from $gN_0=0$ to $gN_0=50$, with increase $dg=0.5$, for different values of $\sigma$. Since we are interested in the properties of the system at low energy, we focus on the nonlinear evolution of the two lowest-energy levels of the double well $m=1,2$, which are the symmetric and antisymmetric states that we denote by $\psi_{\rm s,a}(x)\simeq e^{-(x+x_{\rm min}/2)^2/2\xi^2}\pm e^{-(x-x_{\rm min}/2)^2/2\xi^2}$ respectively at energy $\mu_{\rm s,a}$. In the linear case $gN_0=0$, the symmetric and antisymmetric states are characterized respectively by the absence or presence of a node at $x=0$, i.e., $\psi^{(0)}_{\rm s}(0)\neq0$ while $\psi^{(0)}_{\rm a}(0)=0$. Due to the node in the antisymmetric state, one has $\mu^{(0)}_{\rm a}>\mu^{(0)}_{\rm s}$ (in our numerics, $\mu_{\rm s}^{(0)}\simeq0.364$ and $\mu_{\rm a}^{(0)}\simeq0.522$). The goal now is to study how these two states, and their ordering in energy, change in the presence of the nonlocal nonlinearity.

\subsubsection{Depletion of the condensate density}
The numerically obtained densities in the nonlinear case $|\psi^{(g)}_{\rm s,a}(x)|^2$ are shown in Fig.~\ref{fig:doublewellpotentialwavefunction}, panels (a) and (b). We obtain these states by seeding the Newton-Raphson algorithm using $\psi^{(0)}_{\rm s,a}(x)$ and $\mu_{\rm s,a}^{(0)}$, respectively. We specifically plot $|\psi^{(g)}_{\rm s,a}(x)|^2$ for $gN_0=50$ and for \revthree{$\sigma=0,2,4,6,9$}, to gradually reach the regime of highly nonlocal interaction starting from the local case. From our simulations, in the linear case, we find correctly the symmetric and antisymmetric states (black \revone{dashed} lines). In the nonlinear case, the form of the condensate density drastically depends on the interaction range. In the local case $\sigma=0$ \revthree{(blue solid lines)}, increasing $gN_0$ simply spreads the condensate density within the wells. This result may not come as a surprise, since both condensate peaks tend to reach the Thomas-Fermi limit of very dense photon cloud~\cite{RevModPhys.71.463}, while either overlapping around $x=0$ in the symmetric state, or avoiding the overlap preserving the node in the antisymmetric state.

When the interaction becomes nonlocal, the density at a given point $x$ feels the repulsion from a finite portion of the surrounding density. This fact can drastically affect the low-energy properties of the system, depending on the value of $\sigma$. We identify three distinct nonlocal regimes: (i) A regime where $\sigma$ is smaller than the condensate width $S$, (ii) A regime where $\sigma$ and $S$ are comparable, and (iii) A regime where $\sigma$ becomes larger than $S$. In the first regime (\revthree{blue-gray dash-dotted line for $\sigma=2$} in Fig.~\ref{fig:doublewellpotentialwavefunction}), increasing $\sigma$ qualitatively changes the density profile but does not provide any drastic effect in the low-energy part of the spectrum: The two density lobes reduce their width and repel from each other, effectively renormalizing $x_{\rm min}$. In the symmetric state (Fig.~\ref{fig:doublewellpotentialwavefunction}a), this means that the lobes overlap around $x=0$ starts to decrease. In the antisymmetric case (Fig.~\ref{fig:doublewellpotentialwavefunction}b), instead, the overlap is already suppressed due to the presence of the node at $x=0$, and thus the repulsive nonlocal interaction simply enhances such a repulsion.

A critically different scenario is observed in the second regime (\revthree{dark green dash-dotted line with $\sigma=4$} in Fig.~\ref{fig:doublewellpotentialwavefunction}). We observe that, when starting our calculation from the symmetric state, the nonlocality almost completely suppresses the lobes overlap, therefore inducing a depletion area around $x=0$, and it further pushes the density lobes far apart from each other. The effect is less drastic when starting from the antisymmetric state, because the node already effectively provides a depletion area around $x=0$. This is further enhanced by the nonlocal repulsion, increasing also in this case the lobes mutual distance. Interestingly, due to the induced depletion, the spatial form of the condensate density in the symmetric and antisymmetric cases become very similar, suggesting the onset of a \emph{level inversion} between the symmetric and antisymmetric states. This point is further corroborated below.

Lastly, in the third regime, for $\sigma\gtrsim S$, the effect of the interaction becomes trivial. In this case, the interaction on the scale of the condensate size becomes all-to-all, i.e., the density at each point $x$ interacts with the density of the whole condensate, and thus one can write Eq.~\eqref{eq:grosspitaevskiiequation2} with $K(\mathbf{r},\mathbf{r}')\equiv K_0$, i.e.,
\begin{equation}
\left[-\frac{\hbar^2}{2m}\nabla^2+V(\mathbf{r})\right]\psi(\mathbf{r})=\left(\mu-gN_0K_0\right)\psi(\mathbf{r}) \,\, .
\label{eq:grosspitaevskiiequation2bis1}
\end{equation}
The interaction then simply becomes an overall energy shift that can be included in the definition of $\mu$ and then the system returns to be effectively linear. This is clearly observed in Fig.~\ref{fig:doublewellpotentialwavefunction}, panels (a) and (b), for $\sigma=6$ \revthree{(light green dashed line)}, and especially for $\sigma=9$ (red \revthree{solid} line), where the density becomes equal to the one in the linear case (black dashed line).

\subsubsection{Symmetric-antisymmetric level inversion}
A deeper insight on the system behaviour in the three regimes is provided by Fig.~\ref{fig:doublewellpotentialwavefunction}c. We define the energy gap $\Delta E_{\rm gap}\coloneqq \mu^{(g)}_{\rm a}-\mu^{(g)}_{\rm s}$, and plot it \revthree{in units of $\hbar\omega$} as a function of $gN_0$ for the same values of $\sigma$ used in panels (a) and (b). This allows us to trace the nonlinear evolution of the energies of the symmetric and antisymmetric state: $\Delta E_{\rm gap}>0$ implies that the lowest-energy state of the system is the symmetric one, where the system prefers to keep a nonzero overlap between the density lobes, even if vanishingly small, while $\Delta E_{\rm gap}<0$ implies instead that the lowest-energy state is the antisymmetric one, highlighting the preference for the system to form a node at $x=0$ and thus a \emph{full} depletion area between the lobes. As evident from the figure, \revone{for the} scanned values of $gN_0$, one has $\Delta E_{\rm gap}>0$ for $\sigma=0,2$ and $\sigma=6,9$, corresponding to the first and third regime, respectively \revone{(for $\sigma=6$ we see a monotonically decreasing behaviour of the gap, which may hint a level inversion for $gN_0$ larger than the scanned values)}. In these cases, the ground-state of the system is the symmetric state. Notice that for $\sigma=9$ (red line) one has $\Delta E_{\rm gap}\simeq0.158$, independently of $gN_0$, which is the result in the linear case $\Delta E_{\rm gap}=\mu^{(0)}_{\rm a}-\mu^{(0)}_{\rm s}$ (black dashed line) as expected from Eq.~\eqref{eq:grosspitaevskiiequation2bis1}. Remarkably, in the second regime for \revthree{$\sigma=4$}, $\Delta E_{\rm gap}$ becomes negative for sufficiently large $gN_0$. This is a signature of the level inversion between the symmetric and the antisymmetric state: \revtwo{At this point}, the antisymmetric state, which is the excited state for small $gN_0$, becomes the lowest-energy state.

The analysis in this section points out the highly nontrivial interplay between the strength of the interaction and its range, allowing us to conclude that, in proper regimes, the nonlocal interaction can induce a complete depletion region in the condensate density between the wells. We notice that a double-well potential can be seen as a limit case of a one-dimensional lattice potential, where the two wells identify the two ``boundary'' (or \emph{surface}) sites of the lattice, and the point $x=0$ is the degenerate ``bulk''. As such, a natural question that now arises is how nonlocal nonlinearities affect the condensate density when the potential $V(x)$ describes a one-dimensional lattice, where the presence of both bulk and boundary states can give raise to nontrivial effects.

\begin{figure}[t]
\centering
\includegraphics[width=8cm]{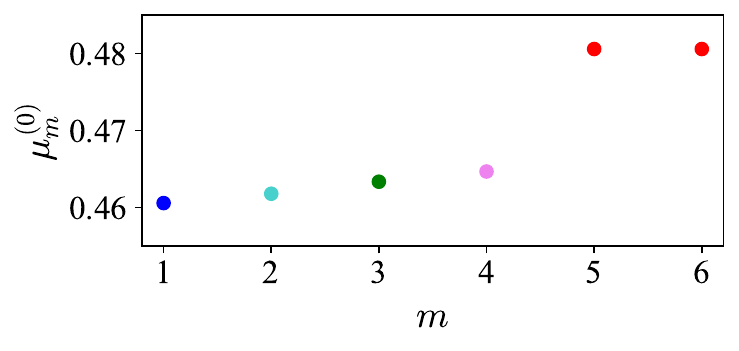}
\caption{First $D_x$ eigenvalues $\{\mu^{(0)}_m\}$ of the linear Hamiltonian $\mathbf{H}$ (in units of $\hbar\omega$) for the lattice with $D_x=6$ in Eq.~\eqref{eq:grosspitaevsiiequationlattice4}. The spectrum consists of $D_x-2$ non-degenerate bulk states (\revthree{$m=1,2,3,4$}) that identify the tight-binding band $\mu_m\simeq\mu_0+J_0\lambda_m$, for some $\mu_0$ and $J_0$, and $\lambda_m$ as in Eq.~\eqref{eq:eigenvaluesgithbingindmodel1}. The two quasi-degenerate surface states (\revthree{$m=5,6$}) appear at higher energy above the bulk states. The respective wavefunctions of these six states are shown in Fig.~\ref{fig:spectrumtightbinding2} of Appendix~\ref{appendix:spectrumtightbindinglimit}.}
\label{fig:spectrumtightbinding1}
\end{figure}

\section{Surface localization of light in lattice potentials}
\label{sec:lattices}
In this section, we extend the discussion in Sec.~\ref{sec:doublewell} by taking as potential $V(x)$ a one-dimensional array of several wells, where a well defines a lattice site. We first review the spectrum of the linear Hamiltonian for the lattice potential, and then study how nonlocal nonlinearities modify the energy landscape, highlighting the onset of surface localization of the condensate density.

\subsection{Model and linear spectrum}
\label{sec:modelandlinearspectrum}
We consider a lattice of $D_x$ sites along $x$, where the distance between two consecutive minima (i.e., the lattice constant) is given by $x_{\rm min}$. Without loss of generality, we take the potential $V(x)$ as an even function of $x$ as
\begin{equation}
V_{\rm L}(x)\!=\!\left\{
\begin{array}{ll}
\!\!\cfrac{m\omega^2}{2}\displaystyle{{\left(\cfrac{x_{\rm min}}{\pi}\right)}^2\!\!\!\cos^2\left(\cfrac{\pi x}{x_{\rm min}}+\phi_x\right)} &  (|x|<x_B)\\\\
\!\!\displaystyle{\cfrac{m\omega^2}{2}{\left(|x|-x_B\right)}^2} & (|x|\geq x_B)
\end{array}
\right. \,\, ,
\label{eq:grosspitaevsiiequationlattice4}
\end{equation}
where the subscript ``L'' stands for ``lattice''.  We denote by $x_B=x_{\rm min}(D_x-1)/2$ the position of the rightmost external minima (i.e., the right boundary of the potential), and $\phi_x=0$ for $D_x$ even and $\phi=\pi/2$ for $D_x$ odd. The different value of $\phi_x$ depending on the parity of $D_x$ ensures the spatial inversion symmetry of $V_{\rm L}(x)$. Notice that $V_{\rm L}(x)$ reduces to $V_{\rm 2W}(x)$ in Eq.~\eqref{eq:equationdoublewellpotential} when $D_x=2$.

Let us consider the case $x_{\rm min}\gg\xi$ (deep lattice, or tight-binding, limit), where the distance between two consecutive lattice wells is much larger than the characteristic size of a lattice well. As in Sec.~\ref{sec:doublewell}, we focus on the low-energy part of the spectrum, specifically the first $D_x$ energy states. The low-energy wavefunctions $\{\psi^{(0)}_m(x)\}$, with $m=1,\ldots,D_x$, consists in general of $D_x$ localized quasi-Gaussian peaks (i.e., the Wannier localized functions), each one centered on a given lattice well. Out of these $D_x$ peaks, two peaks are centered at the two outermost lattice sites (i.e., the boundary, or surface, sites), and the remaining $D_x-2$ peaks occupy the bulk lattice sites. The overlap of the Wannier functions on the bulk sites gives raise to the set of $D_x-2$ energy levels $\{\mu^{(0)}_m\}$ ($m=1,\ldots,D_x-2$), defining the tight-binding lowest energy band. The band becomes flatter and flatter (i.e., the bulk states approach degeneracy) the larger $x_{\rm min}/\xi$. Instead, the two Wannier functions at the boundary sites give raise to two additional, quasi-degenerate energy levels $\mu^{(0)}_{D_x-1}$ and $\mu^{(0)}_{D_x}$, at significantly higher energy compared to the tight-binding band. These two high-energy levels identify the symmetric and antisymmetric localized surface modes, effectively forming a boundary double-well system (see Appendix~\ref{appendix:spectrumtightbindinglimit}).

While the discussion above is valid for general $D_x$, which can be arbitrarily large, in this paper, we focus on a relatively short lattice of $D_x=6$ sites. We make this choice because, \revthree{on one hand, short lattices are relevant to current photon BEC experiment, and on the other hand, they allow} us to systematically investigate the effect of nonlinearities while keeping a reasonable numerical complexity. Always for numerical reasons, we choose $x_{\rm min}/\xi=6$. We identify this choice as a good compromise between being in the sufficiently deep-lattice limit, while avoiding the presence of vanishing small energy gaps in the spectrum that would spoil the numerical convergence of our algorithm. The first $D_x$ eigenvalues, ordered in ascendent order, are shown in Fig.~\ref{fig:spectrumtightbinding1}. Dots with $m=1,2,3,4$ denote bulk states, where the condensate wavefunction is distributed on the four bulk lattice sites. The red dots at $m=5,6$ are the two quasi-degenerate boundary states, where instead the condensate wavefunction focuses on the two boundary sites, see Fig.~\ref{fig:spectrumtightbinding2} of Appendix~\ref{appendix:spectrumtightbindinglimit}.

\subsection{Nonlinearity-induced surface localization}
\label{sec:nonlinearityinducedlocalization}
We now study the effect of a nonlocal nonlinearity on the lattice photon BEC, and provide evidence of the onset of surface localization of the condensate wavefunction. We follows the same numerical scheme used in Sec.~\ref{sec:doublewell} for the double well. Here, we use $M=1000$ and $L_x=150$ (i.e., $\Delta_x=0.15$), and scan the nonlinearity strength from $gN_0=0$ to $gN_0=10$, using a step $dg=0.025$.

\begin{figure}[t]
\centering
\includegraphics[width=4.2cm]{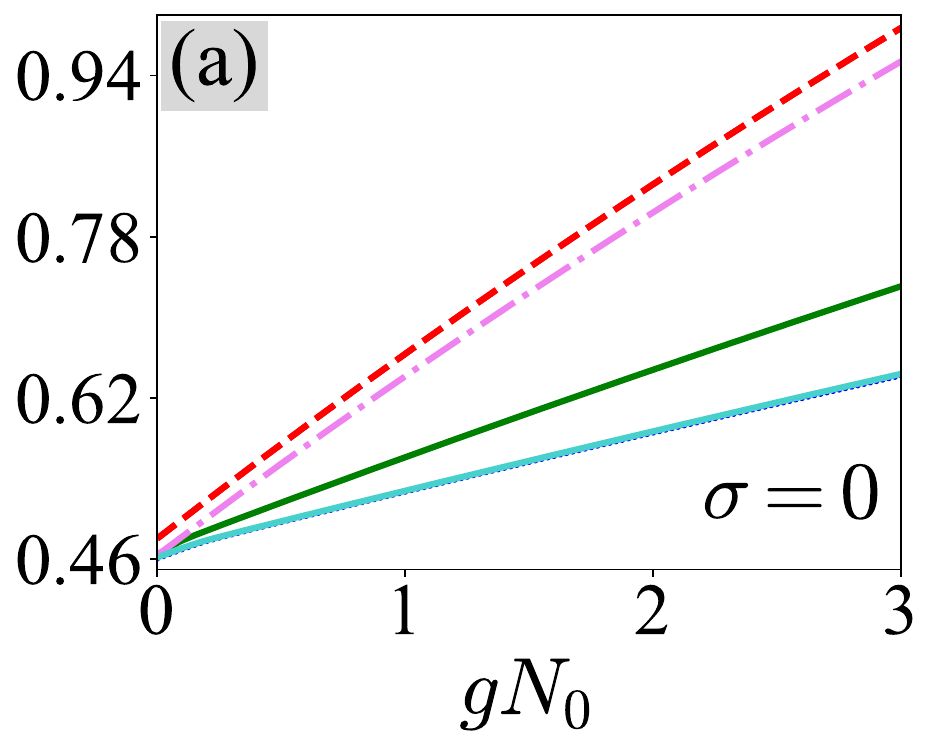}
\includegraphics[width=4.2cm]{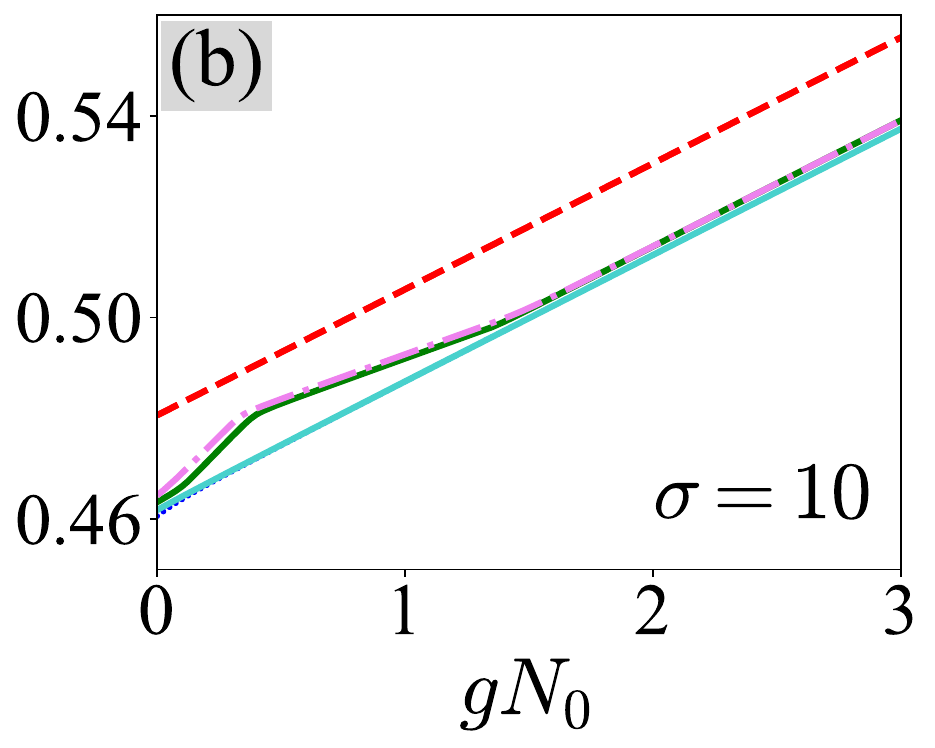}
\includegraphics[width=4.2cm]{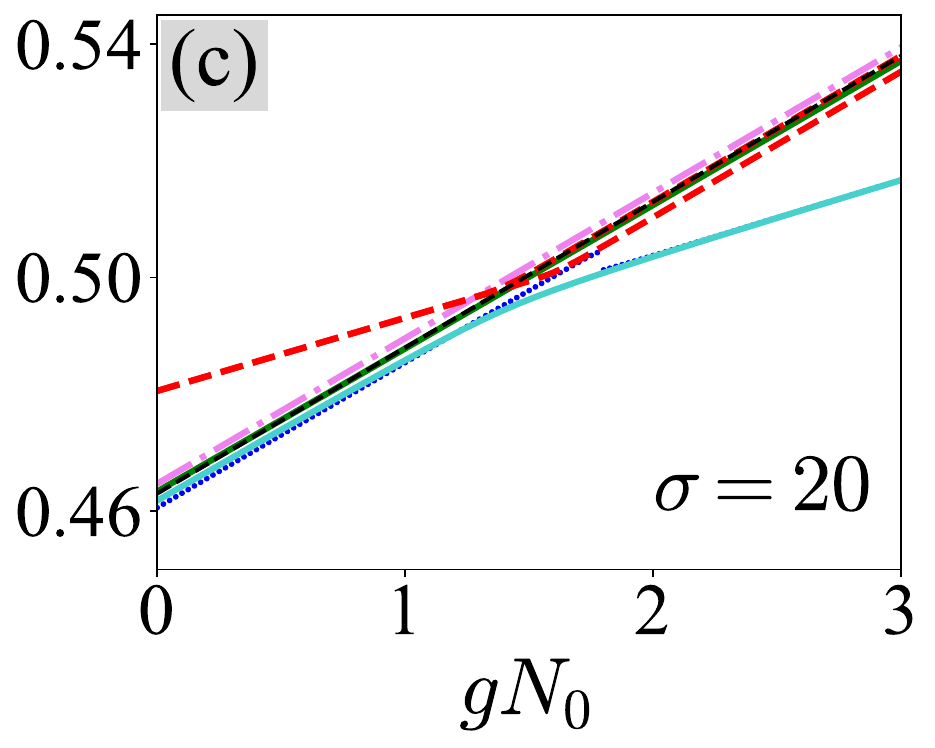}
\includegraphics[width=4.2cm]{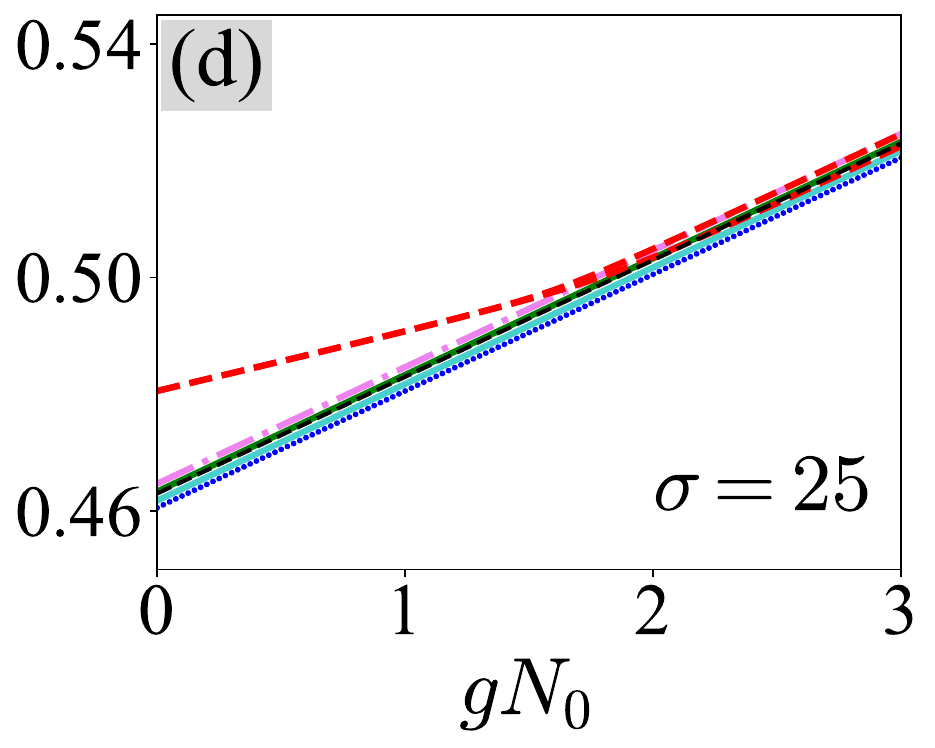}
\caption{Chemical potential $\mu^{(g)}_m$ \revthree{in units of $\hbar\omega$} for $m=1,\dots,D_x$ as a function of $gN_0$ obtained by solving Eq.~\eqref{eq:grosspitaevskiiequation2} for (a) $\sigma=0$, (b) $\sigma=10$, (c) $\sigma=20$, and (d) $\sigma=25$, using as initial guess for the Newton-Raphson calculation $\psi^{(0)}_m(x)$ and $\mu^{(0)}_m$. Color coding as in Fig.~\ref{fig:spectrumtightbinding1}. The data for $m=1$ are plotted as \revthree{blue} dots, and those for $m=5,6$ \revthree{as dashed red lines}. For $\sigma=20$, a level inversion around $gN_0=1.7$ is detected: The boundary state, which is the highest energy state for small $gN_0$, becomes the lowest state for large $gN_0$. The black dashed line in panels (c) and (d) marks the slope $\mu^{(g)}_m-\mu_m^{(0)}=gN_0K_0$ in Eq.~\eqref{eq:grosspitaevskiiequation2bis1} expected in the highly nonlocal limit ($K_0\simeq1/2\sigma$, see Sec.~\ref{sec:doublewell}). The data for $m=1$ and $m=2$ (dark blue dots and light blue \revthree{solid} line, respectively) appear almost overlapped on the scale of the plot. \revtwo{The data are plotted up to $gN_0=3$ for graphical purposes only.}}
\label{fig:wavefunctionandspectrumnonlinearlattice2}
\end{figure}

\begin{figure*}[t]
\centering
\includegraphics[width=4.0cm]{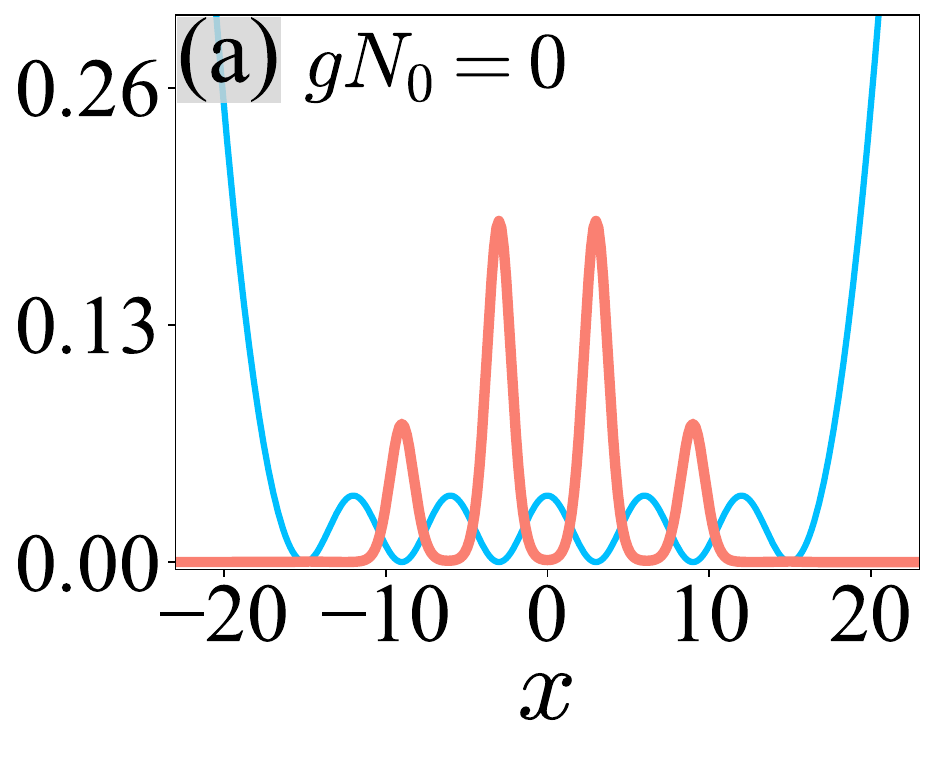}
\includegraphics[width=3.35cm]{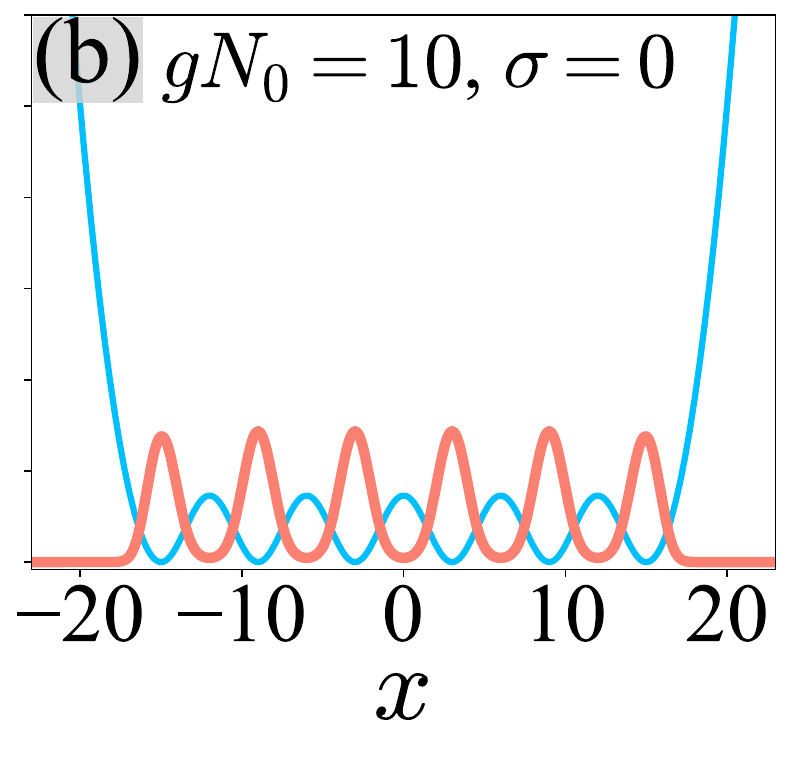}
\includegraphics[width=3.35cm]{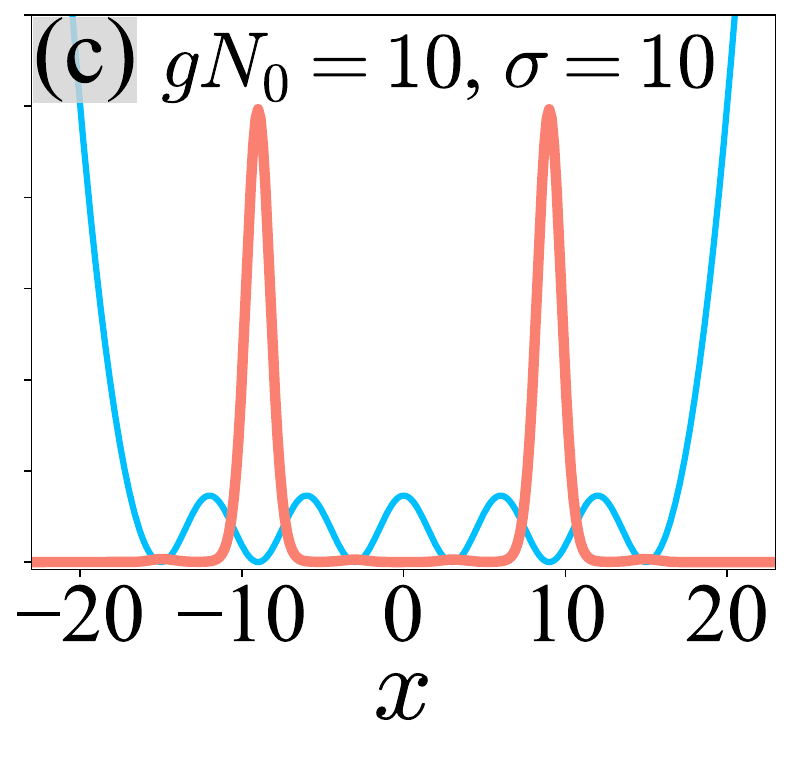}
\includegraphics[width=3.35cm]{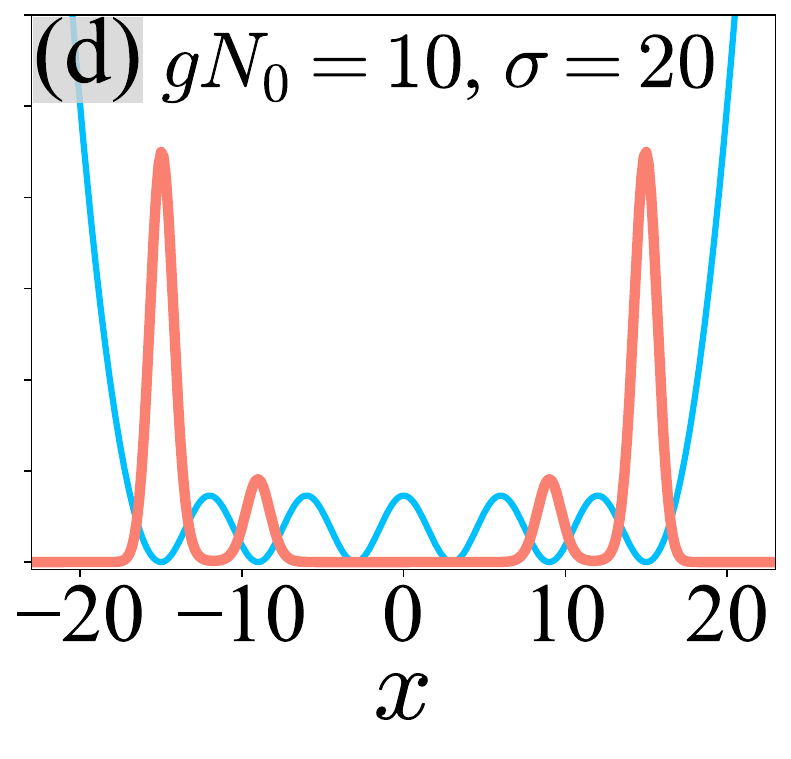}
\includegraphics[width=3.35cm]{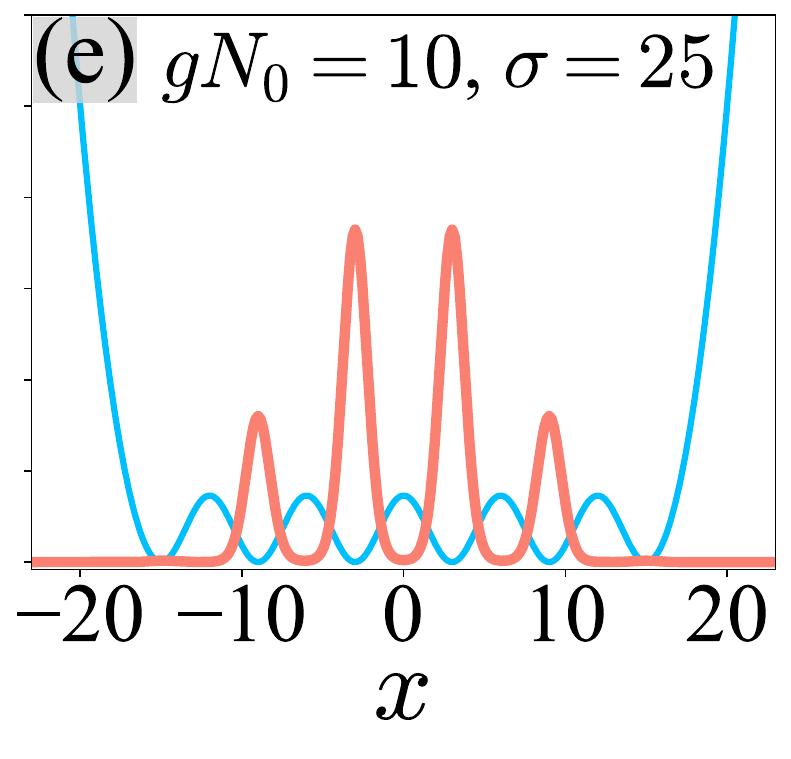}
\caption{Ground-state wavefunction density $|\psi^{(g)}_1(x)|^2$ (pink \revthree{thick} line) and lattice potential $V_{\rm L}(x)$ (blue \revthree{thin} line) in Eq.~\eqref{eq:grosspitaevsiiequationlattice4}, with $\sqrt{m}\omega=1$, $D_x=6$, and $x_{\rm min}=6$. \revthree{Lengths and energies are in units of $\xi$ and $\hbar\omega$, respectively}. The values of the potential are rescaled by a factor $0.02$ for graphical purposes. Panel (a) is the linear solution for $gN_0=0$ (see also Appendix~\ref{appendix:spectrumtightbindinglimit}), where the value of $\sigma$ is irrelevant, while other panels are for $gN_0=10$ and (b) $\sigma=0$, (c) $\sigma=10$, (d) $\sigma=20$, and (e) $\sigma=25$. For $\sigma=20$, the condensate density localizes at the boundary sites.}
\label{fig:wavefunctionandspectrumnonlinearlattice1}
\end{figure*}

The result of our simulation is shown in Fig.~\ref{fig:wavefunctionandspectrumnonlinearlattice2}, where we plot the chemical potential $\mu_m^{(g)}$ for the first $D_x$ states as a function of $gN_0$. We simulate different values of $\sigma$, from the local case $\sigma=0$, where $K(x)=\delta(x)$, to the highly nonlocal case where the interaction range $\sigma$ is sufficiently larger than the spatial extension of the condensate wavefunction \revone{$S\simeq(D_x-1)x_{\rm min}$} along the lattice. For $\sigma>0$, the kernel is as in Eq.~\eqref{eq:equationinteractionkernel}. For a given $m$, the energies $\mu_m^{(g)}$ and respective wavefunctions $\psi^{(g)}_m(x)$ are computed by seeding the Newton-Raphson nonlinear calculation using $\psi^{(0)}_m(x)$ and $\mu^{(0)}_m$ with the same $m$ as initial guess. This allows us to trace the evolution of the first $D_x$ energy levels as the nonlinearity strength and ranges are increased, starting from the linear case in Fig.~\ref{fig:spectrumtightbinding1}. These data are complemented by Fig.~\ref{fig:wavefunctionandspectrumnonlinearlattice1}, where we show the wavefunction density $|\psi^{(g)}_1(x)|^2$ for specific cases in Fig.~\ref{fig:wavefunctionandspectrumnonlinearlattice2}. The picture emerging from our numerical data can be summarized as in the following sections.

\subsubsection{Local interaction}
For a local interaction ($\sigma=0$), the first $D_x$ eigenvalues increase with $gN_0$ preserving their ordering, i.e., no level inversion occurs (see Fig.~\ref{fig:wavefunctionandspectrumnonlinearlattice2}a). In particular, the ground state (blue dots) for $gN_0=0$ continuously evolves into the ground state for large $gN_0$. This point is further established by looking at the condensate density in Fig.~\ref{fig:wavefunctionandspectrumnonlinearlattice1}, panels (a) and (b): Starting from the linear solution (see Appendix~\ref{appendix:spectrumtightbindinglimit}), increasing the interaction strength simply spreads the condensate density all over the lattice, tending to a configuration where all the $D_x$ lattice wells are populated and the peaks of the density have the same height on each well. The physical reason for this fact is readily understood. In our system, there are two elements that play an opposite role in terms of energy minimization: On one hand, the lattice potential boundaries act as much as possible to confine the wavefunction both avoiding populating the boundary sites and maximizing the occupation of the central bulk sites, while on the other hand, the local repulsive interaction tends to avoid such an imbalanced occupation, therefore inducing a more and more uniform peak density the larger $gN_0$ along the lattice wells. As such, any form of condensate density localization is energetically unfavorable.

\subsubsection{Non-local interaction}
As the interaction range increases ($\sigma>0$), the system displays a richer phenomenology. We here report the numerical results for $\sigma=10,20,25$. This choice of the values of $\sigma$ is motivated by the fact that we are interested in studying the system where nonlocal effects play a predominant role, i.e., $\sigma\gtrsim x_{\rm min}$, following Sec.~\ref{sec:doublewell}. The nonlinear chemical potentials as a function of $gN_0$ are shown in Fig.~\ref{fig:wavefunctionandspectrumnonlinearlattice2}, panels (b)-(d), and the respective ground-state condensate density is reported in Fig.~\ref{fig:wavefunctionandspectrumnonlinearlattice1}, panels (c)-(e).

We first comment the data for $\sigma=10$ in Fig.~\ref{fig:wavefunctionandspectrumnonlinearlattice2}b, where the interaction range involves several condensate peaks, but it is still smaller than the condensate size $S$. The first $D_x$ nonlinear eigenvalues evolve with $gN_0$ starting from their respective linear values by preserving their ordering, as in the local case. We observe that the bulk eigenvalues for $m=1,2$ and $m=3,4$, after an initial separation, become very close in energy, however, they are always well separated in energy from the boundary eigenvalues for $m=5,6$ (red \revthree{dashed} lines). The ground-state density shown in Fig.~\ref{fig:wavefunctionandspectrumnonlinearlattice1}c displays two peaks at the sites close to the boundary ones, and the two central bulk sites are almost empty. This effect can be seen simply as a redistribution of the condensate density within the bulk sites, in order for the system to minimize the effect of the nonlocal interaction, while always keeping the boundary sites unoccupied.

As the interaction range is further increased and becomes comparable with the condensate size $S$, a different scenario arises. The key result of our paper emerges for $\sigma=20$, in Fig.~\ref{fig:wavefunctionandspectrumnonlinearlattice2}c. For small $gN_0$, we see that the increase of the $D_x-2$ bulk eigenvalues with $gN_0$ is well captured by the linear behaviour $\mu^{(g)}_m-\mu^{(0)}_m\simeq gN_0K_0$, with $K_0\simeq1/2\sigma$, expected from the highly nonlocal approximation in Eq.~\eqref{eq:grosspitaevskiiequation2bis1} (black dashed line). Instead, the boundary eigenvalues ($m=5,6$, red lines), in striking difference with the previous cases, increase with a smaller slope, and eventually evolve into a bulk state around $gN_0=1.7$. Around this value, the ground-state and first-excited state eigenvalue ($m=1$ and $m=2$ corresponding to dark blue dots and light blue line in the figure, which are almost overlapped), invert with the boundary eigenvalue, and evolve into a (nonlinear) boundary state. \revone{Here, specifically, the state with $m=2$, which has a node in $x=0$, evolves into the lowest-energy state}. As such, the boundary state that is the highest state in energy in the lower band for $gN_0\lesssim1.7$ becomes the ground state for $gN_0\gtrsim1.7$. This fact is also clearly seen from the form of the condensate density in Fig.~\ref{fig:wavefunctionandspectrumnonlinearlattice1}, comparing panel (a), which is the linear solution for $gN_0=0$, with panel (d). After the level inversion, the increase of the eigenvalues for $m=3,4,5,6$ is again well captured by the highly nonlocal approximation.

By further increasing $\sigma$ towards the highly nonlocal regime, this level inversion, and thus surface localization, is not detected within the scanned range of $gN_0$. This is seen in Fig.~\ref{fig:wavefunctionandspectrumnonlinearlattice2}d for $\sigma=25$. The boundary state evolves into a bulk state in a similar way as in panel (c), but all other eigenvalues evolve with $gN_0$ according to the highly nonlocal approximation in Eq.~\eqref{eq:grosspitaevskiiequation2bis1}. Here, the ground-state density preserves its shape throughout the nonlinear evolution, as can be seen by comparing panels (a) and (e) of Fig.~\ref{fig:wavefunctionandspectrumnonlinearlattice1}, in agreement with the fact that the system is entering the all-to-all interaction regime and remains effectively linear.

The fact that a surface localized state in this intermediate nonlocal regime becomes the most energetically favorable state can be intuitively explained as follows. As said in the local case, for small interaction strength, the most energetically favorable configuration is always the one having a condensate density focused mostly on the central lattice sites, due to the presence of the potential boundaries. When the interaction strength becomes sufficiently large, and the interaction range is comparable with the size of the condensate density, despite the presence of the lattice boundaries, it becomes more energetically favorable for the system to minimize as much as possible the interaction energy by splitting the condensate density focusing it at the lattice boundaries.

More rigorously, the interaction term in the intermediate regime behaves as a correction to the lattice potential $V_{\rm L}(x)$, raising the energy of the bulk sites with respect to the boundary sites. This can be seen as follows. If we call $x_j$ the position of the $j$-th minimum of the lattice potential, one can approximate in the limit $\sigma\gg\xi$ the density as $\left|\psi(x)\right|^2\simeq\sum_j\left|\psi(x_j)\right|^2\delta(x-x_j)$. The interaction term in the Gross-Pitaevskii equation can then be approximated as $\int dx'\,K\left(x-x'\right)\left|\psi\left(x'\right)\right|^2\simeq\sum_jK(x-x_j)\left|\psi(x_j)\right|^2$. This term behaves as a correction to the lattice potential $V_{\rm L}(x)$, adding a local nonlinear chemical potential $\mu_{\rm NL}(x)=gN_0\sum_jK(x-x_j)\left|\psi(x_j)\right|^2$. For $gN_0>0$, $\mu_{\rm NL}(x)$ is maximal at the center of the lattice, and minimum at the boundaries. This can be seen by observing that, for $x$ close to the lattice center, the sum over $j$ will symmetrically include sites both to the left and to the right of $x$ [recall that $K(x)$ is a symmetric function of $x$]. As $x$ gradually approaches the lattice boundaries, the summation decreases because it includes an increasing number of spatial points both towards the surface, where $\left|\psi(x_j)\right|^2\simeq0$, and far from it, where instead $K(x-x_j)\simeq0$, yielding a zero contribution to the overall sum. As a result, the minima of the overall effective lattice potential $V_{\rm eff}(x)\equiv V_{\rm L}(x)+\mu_{\rm NL}(x)$ will be at lower energy at the boundaries with respect to the lattice center, hence favoring the focusing of condensate density close to lattice boundaries.

\section{Conclusions and perspectives}
\label{sec:conclusions}
In conclusion, we theoretically studied the effect of a repulsive (defocusing) thermo-optical nonlocal nonlinearity in Bose-Einstein condensate (BEC) of photons. We first studied the case of photons trapped in the double-well potential experimentally realized in Ref.~\cite{doi:10.1126/science.aay1334}, in the presence of a \revtwo{strong} nonlocal interaction with tunable range. We performed our analysis by developing a numerical code inspired by \revtwo{the} Newton-Raphson method, which allowed us to solve the time-independent Gross-Pitaevskii equation with arbitrary potential and interaction form. We observed that, while a local interaction induced a spreading of the condensate density throughout the double well, a sufficiently large but finite nonlocality favored instead the emergence of a full depletion area between the two wells, tending to focus the condensate density at the potential boundaries.

We then extended our analysis to a one-dimensional small lattice of six sites. This allowed us to generalize our previous observations for the double well, which is a minimal extension of a lattice with two sites only, to a more complex case where a distinction between bulk and boundary (or surface) sites could be made. We focused on the nonlinear evolution of the first low-energy eigenvalues forming the lowest energy band for different values of the interaction range. Our key results was that, when the interaction range was comparable with the size of the lattice, and the interaction was sufficiently strong, a level inversion between eigenvalues took place. In particular, the state with the condensate wavefunction localized at the potential boundaries, which is the highest eigenvalue of the lowest band for a weak or short-range interaction, becomes the ground-state energy level for strong interaction. This fact signaled the onset of surface localization, where the condensate density is focused at the system boundaries. \revone{In our paper, we modeled the photon-photon thermo-optical interaction as a regularized box potential. In future, it would be interesting to study the emergence of surface localization using smoother forms of the interaction kernel}\revthree{, as well as the extension of our results on surface localization to large lattices with several sites.}

Our results point out the highly nontrivial interplay between potential boundaries and interaction within a one-dimensional array of potential wells. An intriguing question opened by our work is how long-range interactions affect the photon BEC when several one-dimensional lattices are coupled together, with the inclusion of a complex coupling between photons in different arrays to mimic the presence of a synthetic magnetic (or gauge) flux in the system (i.e., a bosonic flux ladder of photons). A possible way to introduce an effective magnetic field in our setup, thereby implementing a photonic two-leg flux ladder, can be making two separated structured microcavities, each one realizing a one-dimensional array of photon BEC, and couple them with a time-modulated coupling constant, with a proper distribution of the modulation phases in space~\cite{fangeffectivemagneticfieldphotons2012}. \revtwo{The} interaction range is determined by the degree of nonlocality of the thermo-optical response of the microcavity, which can be controlled by adding proper thermo-responsive dopants to the system~\cite{PhysRevA.90.043853,PhysRevResearch.3.023167}. \revtwo{In the present experiments using dye molecules~\cite{weitz2010becphotons1,weitz2010becphotons2,doi:10.1126/science.aay1334}, the pumping is not performed continuously but rather with $500\,{\rm ns}$ long pulses. One one hand, this time is much longer than the time scale of the absorption and reemission processes by the dye molecules (approximately between $10\,{\rm ps}$ and $100\,{\rm ps}$), inducing effective thermalization of the photon gas. On the other hand, this pulse length is shorter than the expected equilibration time for heat transport to reach a steady state within the dye solution. True continuous wave operation is expected to be possible for instance using semiconductor materials for photon thermalization~\cite{Barland21}}.

\revtwo{Our} work paves the way towards the simulation of condensed-matter systems using photon BECs (i.e., synthetic matter of light), thus envisioning the intriguing possibility to study of exotic, possibly topological phases of matter emerging from the interplay between low dimensionality, effective gauge fields, and strong interactions with tunable range.

\section*{Acknowledgements}
\revtwo{We thank Julian Schmitt and Martin Weitz for fruitful discussions and comments on this paper.} We acknowledge funding from H2020 \revthree{PhoQuS} project (Grant No. 820392). M.~C.~S. and C.~C. acknowledge funding from Sapienza Ricerca, PRIN PELM (20177PSCKT), QuantERA ERA-NET Co-fund (Grant No. 731473, project QUOMPLEX). F.~V. acknowledges funding by the DFG within SFB/TR185 (Project No. 277625399).

\appendix

\section{Fitting the double-well potential}
\label{appendix:detailsonthefitdoublewellpotential}
In this appendix, we provide details on the way we fit the experimental data of the double-well potential from Ref.~\cite{doi:10.1126/science.aay1334} with the function in Eq.~\eqref{eq:equationdoublewellpotential}. We estimate the value of the trapping frequency $\omega$ from the value of the height of the potential barrier $\Delta E^{{\rm (th.)}}\coloneqq V_{\rm 2W}(0)-V_{\rm 2W}(\pm x_{\rm min}/2)$, which from Eq.~\eqref{eq:equationdoublewellpotential} is
\begin{equation}
\Delta E^{{\rm (th.)}}=\frac{m\omega^2}{2}{\left(\frac{x_{\rm min}}{\pi}\right)}^2 \,\, .
\label{eq:potentialbarriervalue1}
\end{equation}
We impose that the value in Eq.~\eqref{eq:potentialbarriervalue1} matches the experimental value of the potential barrier $\Delta E^{\rm(exp.)}$, which from the data in Fig.~\ref{fig:potentialdoublewellweitz} is $\Delta E^{\rm(exp.)}\simeq0.109\,{\rm THz}$. By inverting the relation in Eq.~\eqref{eq:potentialbarriervalue1} and by imposing $\Delta E^{{\rm (th.)}}\equiv\Delta E^{\rm(exp.)}$, one has $\sqrt{m}\omega=(\pi/x_{\rm min})\sqrt{2\Delta E^{\rm(exp.)}}$. By using $x_{\rm min}\simeq12.5\,{\rm \mu m}$, which is the experimental value of the distance between the two minima of the double-well potential, we find $\sqrt{m}\omega\simeq0.117346\,{\rm \sqrt{THz}/\mu m}$. The trapping frequency $\omega$ can be then estimated by assuming a photon mass $m\simeq7.76271\times10^{-36}\,{\rm kg}$~\cite{doi:10.1126/science.aay1334}, and one finds $\omega\simeq1.08415\times10^{12}\,{\rm rad/s}$ (approximately $0.173\,{\rm THz}$). This yields the characteristic length scale for each local harmonic oscillator $\xi=\sqrt{\hbar/m\omega}\simeq3.53987\,{\rm \mu m}$, which is the characteristic length scale in our simulations. In units of $\xi$, the distance between the two minima of the double well is $x_{\rm min}/\xi\simeq 3.53121$. In rescaled units, we then simulate Eq.~\eqref{eq:grosspitaevskiiequation2} setting $\hbar/\sqrt{m}=1$, $\sqrt{m}\omega=1$, and $x_{\rm min}=3.53121$, i.e., giving lengths and energies in units of $\xi$ and $\hbar\omega$, respectively. As a final annotation, we mention that, to fit the experimental data in Fig.~\ref{fig:potentialdoublewellweitz}, the potential in Eq.~\eqref{eq:equationdoublewellpotential} was shifted by $V_{\rm min}=0.12\,{\rm THz}$, so that $V_{\rm 2W}(\pm x_{\rm min}/2)=V_{\rm min}$, but this offset just adds a constant shift to the chemical potential $\mu$ in Eq.~\eqref{eq:grosspitaevskiiequation2}, and we then take $V_{\rm min}=0$ in our simulations.

\section{Details on the numerical simulation}
\label{appendix:detailsonthenumericalsimulations}
In this appendix, we discuss in detail the numerical method used to solve the Gross-Pitaevskii equation
\begin{eqnarray}
&&\left[-\frac{1}{2}\frac{d^2}{dx^2}\!+\!V(x)\!+\!gN_0\!\int_{-\infty}^{\infty} dx'\,K(x\!-\!x'){|\psi(x')|}^2\right]\!\psi(x)\nonumber\\
&&=\mu\,\psi(x) \,\, ,
\label{eq:numericalmethods1}
\end{eqnarray}
\revtwo{in dimensionless units (see main text).} In Eq.~\eqref{eq:numericalmethods1}, we take the wavefunction and the nonlinear kernel normalized to unity, $\int_{-\infty}^{\infty} dx\,{|\psi(x)|}^2=1$ and $\int_{-\infty}^{\infty}dx\,K(x)=1$. In addition, the potential $V(x)$ and nonlinear interaction kernel $K(x)$ are real functions of $x$, so both the wavefunction and chemical potential $\mu$ can be taken real. In our simulations, we always consider cases where $V(x)$ and $K(x)$ are such that the wavefunction $\psi(x)$ is contained within a segment $I$ of length $S$ of the real axis, i.e., $|\psi(x)|\simeq0$ for all $x\notin I$. This allows us to study Eq.~\eqref{eq:numericalmethods1} limiting ourselves to $x$ defined within a finite domain of size $L_x>S$, which we take by convention symmetric with respect to $x=0$, i.e., $-L_x/2\leq x<L_x/2$. We then discretize the $x$ variable as a linear grid of $M$ points equally spaced by $\Delta_x=L_x/M$, i.e., $x_j=-L_x/2+j\Delta_x$, where $j=0,\ldots,M-1$. The wavefunction $\psi(x)$ is then represented by a vector with $M$ entries $\vec\psi=(\psi(x_0),\psi(x_1),\ldots,\psi(x_{M-1}))$. In the following, we will denote by $\psi_j$ the $j$-th entry of $\vec\psi$, i.e., $\psi_j\equiv\psi(x_j)$. It is understood that $L_x$ is sufficiently larger than $S$ to yield $|\psi_j|=0$, for all $j$ with $|x_j|>L_x/2$.

The derivative operator in Eq.~\eqref{eq:numericalmethods1}, acting on the wavefunction $\vec\psi$, is discretized using the pseudo-spectral representation~\cite{boyd2013chebyshev}. By defining the matrix $\mathbf{\Theta}$ as
\begin{equation}
\mathbf{\Theta^1}|_{jk}=(1-\delta_{jk})\frac{{(-1)}^{j+k}}{2}\cot\left(\frac{\theta_j-\theta_k}{2}\right)  \,\, .
\label{eq:grosspitaevskiiequation9bis}
\end{equation}
where $\delta_{jk}$ is the Kr\"onecker delta and $\theta_j=2\pi j/M$, we obtain a matrix representation $\mathbf{D^1}$ of the first-derivative operator $d/dx$ as
\begin{equation}
\mathbf{D^1}=\frac{2\pi}{L_x}\,\mathbf{\Theta^1} \,\, ,
\label{eq:numericalmethods2}
\end{equation}
and consequently the matrix representation $\mathbf{D^2}$ of the second-derivative operator $d^2/dx^2$ reads $\mathbf{D^2}=\mathbf{D^1}\cdot\mathbf{D^1}$, where ``$\cdot$'' denotes the matrix dot product. The matrix representation of the potential $V(x)$ and nonlocal nonlinear convolution $F[\psi(x),x]=\int_{-\infty}^{\infty} dx'\,K(x-x'){|\psi(x')|}^2$ are straightforward, since both terms represent multiplicative operators in $x$ space. Then, $V(x)$ is represented by the diagonal matrix $\mathbf{V}$ with entries $\mathbf{V}|_{jk}=\delta_{jk}V(x_j)$, and the convolution is represented by a diagonal matrix $\mathbf{F}[\vec\psi]$ with entries $\mathbf{F}|_{jk}=\delta_{jk}\,\Delta_x\sum_{p=0}^{M-1}K_{jp}\psi^2_p$. Therefore, the discrete version of Eq.~\eqref{eq:numericalmethods1} reads
\begin{equation}
\left(-\frac{1}{2}\mathbf{D^2}+\mathbf{V}+gN_0\,\mathbf{F}[\vec\psi\,]\right)\vec\psi=\mu\vec\psi \,\, .
\label{eq:numericalmethods2}
\end{equation}
Notice that, in the linear case for $gN_0=0$, Eq.~\eqref{eq:numericalmethods2} reduces to the Schr\"odinger equation for a particle in a potential $V(x)$, which can be solved by exact diagonalzation of the Hamiltonian matrix $\mathbf{H}=-\mathbf{D^2}/2+\mathbf{V}$. To solve Eq.~\eqref{eq:numericalmethods2} for $gN_0\neq0$, we resort to an iteration scheme based on the Newton-Raphson method~\cite{ortega1970iterative,press2002numerical}. The basic idea of this method is to find the unknown wavefunction $\vec\psi$ and chemical potential $\mu$ for a given value of the nonlinearity strength $gN_0\equiv g_0$, which we denote by $\vec\psi^{(g_0)}$ and $\mu^{(g_0)}$ for simplicity, starting from the knowledge of the wavefunction and chemical potential for a smaller value of the nonlinearity strength $g_0-dg$, i.e., $\vec\psi^{(g_0-dg)}$ and $\mu^{(g_0-dg)}$, where ideally $dg$ is infinitesimally small.

If we denote by $\vec\chi\equiv\vec\psi^{(g_0-dg)}$ the known wavefunction, we expand the unknown wavefunction $\vec\psi^{(g_0)}$ as $\vec\psi^{(g_0)}=\vec\chi+\vec\varphi$, where $\vec\varphi$ is a small correction to $\vec\chi$ such that $|\varphi_j|\ll|\chi_j|$, for all $j$, which has to be found by solving Eq.~\eqref{eq:numericalmethods2}. By plugging the expansion $\vec\psi^{(g_0)}=\vec\chi+\vec\varphi$ into Eq.~\eqref{eq:numericalmethods2}, we can reduce the nonlinear equation~\eqref{eq:numericalmethods2} to a \emph{linear} system for the correction $\vec\varphi$. Let us now find the linear equation for $\varphi_j$. By keeping only first-order terms in $\varphi_j$, one has $\psi_p^2\psi_j\simeq\chi^2_p\chi_j+\chi^2_p\varphi_j+2\chi_p\chi_j\varphi_p$ and one can write Eq.~\eqref{eq:numericalmethods2} component-wise as
\begin{eqnarray}
&&\mu\chi_j\!+\!\mu\varphi_j=\sum_hH_{jh}\chi_h+\sum_hH_{jh}\varphi_h\!+\!g_0\Delta_x\chi_j\sum_pK_{jp}\chi^2_p\nonumber\\
&&\hspace{0.5cm}+2g_0\Delta_x\chi_j\sum_pK_{jp}\chi_p\varphi_p+g_0\Delta_x\varphi_j\sum_pK_{jp}\chi^2_p \,\, ,
\label{eq:grosspitaevskiiequation11}
\end{eqnarray}
where $h,p=0,\ldots,M-1$. The goal now is to write from Eq.~\eqref{eq:grosspitaevskiiequation11} a linear system of the form $\mathbf{A}\vec{\varphi}=\vec{v}$ for the correction $\vec{\varphi}$, for a matrix $\mathbf{A}$ and vector $\vec{v}$ to be determined from Eq.~\eqref{eq:grosspitaevskiiequation11}. In this way, one can find the correction as $\vec{\varphi}=\mathbf{A}^{-1}\vec{v}$. It can be useful to rename the dummy indexes $p$ as $h$ in the summations involving $\varphi_p$, rename $\varphi_j=\sum_h\delta_{jh}\varphi_h$, and group terms in Eq.~\eqref{eq:grosspitaevskiiequation11} as
\begin{eqnarray}
&&\sum_h\!\left(\!H_{jh}\!+\!2g_0\Delta_xK_{jh}\chi_j\chi_h\!+\!\delta_{jh}g_0\Delta_x\!\!\sum_p\!K_{jp}\chi^2_p\!-\!\mu\delta_{jh}\!\right)\!\varphi_h\nonumber\\
&&=\mu\chi_j-\sum_hH_{jh}\chi_h-g_0\Delta_x\chi_j\sum_pK_{jp}\chi^2_p \,\, ,
\label{eq:grosspitaevskiiequation13}
\end{eqnarray}
which suggests that one can define the matrix $\mathbf{A}$ and the vector $\vec{v}$ as
\begin{eqnarray}
\mathbf{A}|_{jh}&=&H_{jh}\!+\!2g_0\Delta_xK_{jh}\chi_j\chi_h\!+\!\delta_{jh}g_0\Delta_x\sum_pK_{jp}\chi^2_p\!-\!\mu\delta_{jh}\nonumber\\
v_j&=&\mu\chi_j-\sum_hH_{jh}\chi_h-g_0\Delta_x\chi_j\sum_pK_{jp}\chi^2_p \,\, .
\label{eq:grosspitaevskiiequation14}
\end{eqnarray}
Once $\vec\varphi$ is determined as $\vec{\varphi}=\mathbf{A}^{-1}\vec{v}$, the eigenvalue $\mu\equiv\mu^{(g_0)}$ in Eq.~\eqref{eq:grosspitaevskiiequation13} can be found from the new wavefunction $\vec\psi^{(g_0)}=\vec\chi+\vec\varphi$ simply as
\begin{equation}
\mu^{(g_0)}\simeq\Delta_x\sum_j\psi_j^{(g_0)}\Phi_j \,\, ,
\label{eq:grosspitaevskiiequation16}
\end{equation}
where
\begin{equation}
\Phi_j=\sum_hH_{jh}\psi^{(g_0)}_h+g_0\Delta_x\sum_pK_{jp}\left(\psi^{(g_0)}_p\right)^2\psi^{(g_0)}_j \,\, .
\label{eq:grosspitaevskiiequation16bis0}
\end{equation}
In this method, since the wavefunction and the chemical potential for a nonlinear strength $g_0-dg$ have to be known in order to find those for a nonlinear strength $g_0$, a possible way to determine $\psi_j^{(g_0)}$ and $\mu^{(g_0)}$ for the desired (final) value $g_0$ of the nonlinear strength can be seeding the calculation in Eq.~\eqref{eq:grosspitaevskiiequation13} using as a starting point the wavefunction $\vec\psi^{(0)}$ and $\mu^{(0)}$ computed in the absence of nonlinearities ($g_0=0$), i.e., by solving the Schr\"odinger equation by exact diagonalization of the Hamiltonian matrix $\mathbf{H}$ in Eq.~\eqref{eq:numericalmethods2}. Basically, one first finds $\vec\psi^{(dg)}$ and $\mu^{(dg)}$ seeding Eq.~\eqref{eq:grosspitaevskiiequation13} with $\vec\psi^{(0)}$ and $\mu^{(0)}$. One then repeats the calculation finding $\vec\psi^{(2dg)}$ and $\mu^{(2dg)}$ seeding Eq.~\eqref{eq:grosspitaevskiiequation13} with the previously determined $\vec\psi^{(dg)}$ and $\mu^{(dg)}$, and so on repeating the procedure until $\vec\psi^{(g_0)}$ and $\mu^{(g_0)}$ for the desired value of $g_0$ are reached.

Notice that, in an actual numerical context, where $dg$ is small but anyhow finite, one can still rely on this linearization but one has to perform additional convergence iterations, labelled by $n$, for a given value $g$ of the nonlinear strength in order to refine the value of $\vec\varphi$. In practice, the numerical procedure that we employ to find the solution of Eq.~\eqref{eq:numericalmethods2} for a target value $g_0$ of the nonlinearity strength can be summarized in the following steps:

\vspace{0.2cm}
\underline{1. Solution of the linear problem}. First, one solves the Schr\"odinger equation by exact diagonalization of $\mathbf{H}$ in Eq.~\eqref{eq:numericalmethods2}, and selects the desired state from the linear spectrum. This yields the wavefunction and chemical potential $\vec\psi^{(0)}$ and $\mu^{(0)}$ to seed the nonlinear calculation.

\vspace{0.2cm}
\underline{2. Newton-Raphson nonlinear steps} with gradual increase of the nonlinearity strength. Here, one chooses a small step $dg$ and solves Eq.~\eqref{eq:grosspitaevskiiequation13} as explained before. In order to refine the solution, one performs a number of convergence steps by repeatedly solving Eq.~\eqref{eq:grosspitaevskiiequation13} for a given value $g$. For a given convergence step $n$, one computes $\vec\psi^{(g)}_n$ and $\mu^{(g)}_n$ by finding $\vec\varphi_n$ seeding Eq.~\eqref{eq:grosspitaevskiiequation13} with $\vec\chi\equiv\vec\psi^{(g)}_{n-1}$ and $\mu\equiv\mu^{(g)}_{n-1}$ found from the previous convergence step. To enhance the numerical convergence of the algorithm, $\vec\psi^{(g)}_n$ is updated using a relaxation factor $\gamma<1$~\cite{ortega1970iterative}, i.e., $\vec\psi^{(g)}_n=\vec\chi+\gamma\vec\varphi$, and the chemical potential $\mu^{(g)}_n$ is found as in Eq.~\eqref{eq:grosspitaevskiiequation16}. The wavefunction $\vec\psi^{(g)}_n$ is normalized to unity at every convergence step. During the convergence steps, one can define the maximum numerical error as
\begin{equation}
{\rm MaxErr}(n)\coloneqq{\rm max}_j\left|\left|\psi^{(g)}_{n,j}\right|-\left|\psi^{(g)}_{n-1,j}\right|\right| \,\, .
\end{equation}
By fixing a numerical tolerance $\epsilon\ll1$, the convergence steps are repeated until the condition ${\rm MaxErr}(\bar n)<\epsilon$ is found after a number $\bar n$ of steps. The resulting wavefunction $\vec\psi^{(g)}\equiv\vec\psi^{(g)}_{\bar n}$ and chemical potential $\mu^{(g)}\equiv\mu^{(g)}_{\bar n}$ at the end of the convergence steps are then taken as solution of Eq.~\eqref{eq:numericalmethods2} with nonlinearity strength $g$. These values are used to seed the nonlinear calculation for a nonlinearity strength $g+dg$, and the whole procedure is repeated until the value $g_0$ is reached.

\begin{figure*}[t]
\includegraphics[width=2.8cm]{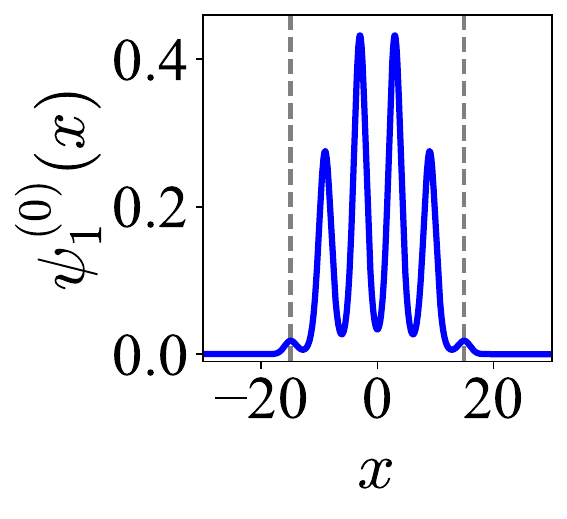}
\includegraphics[width=2.95cm]{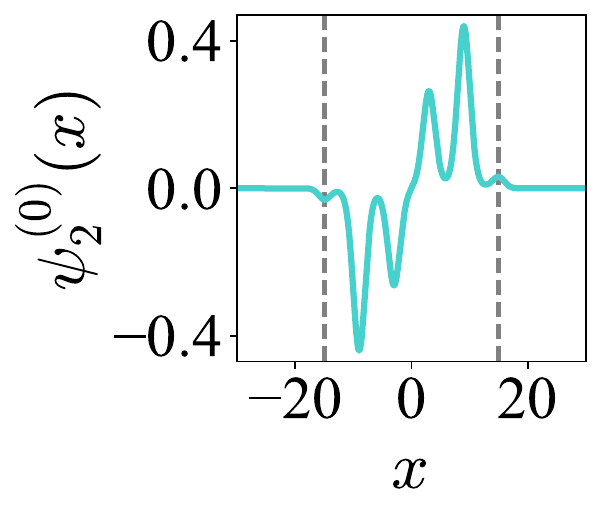}
\includegraphics[width=2.95cm]{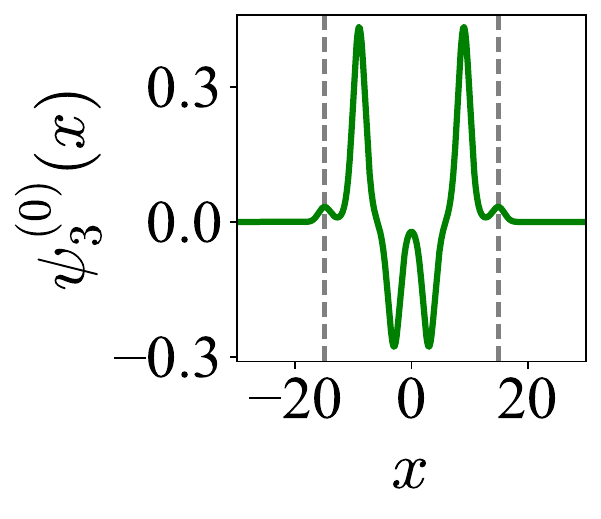}
\includegraphics[width=2.95cm]{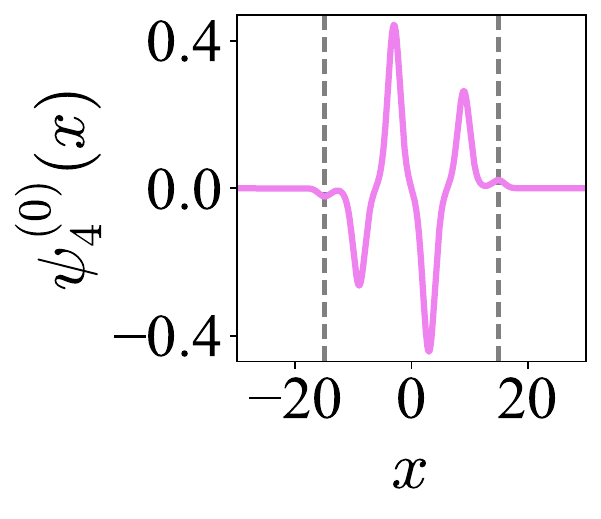}
\includegraphics[width=2.73cm]{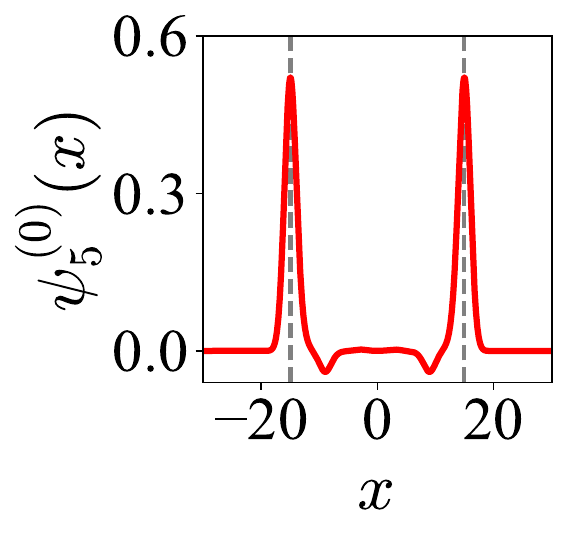}
\includegraphics[width=2.9cm]{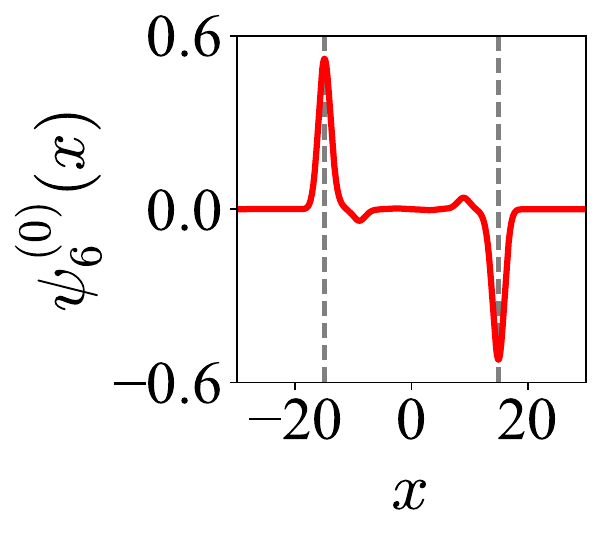}
\caption{Wavefunctions $\{\psi^{(0)}_m(x)\}$ for the first $D_x=6$ levels in Fig.~\ref{fig:spectrumtightbinding1} with the came color coding. The first four panels refer to the four bulk states for $m=1,2,3,4$, and the last two are the two boundary states for $m=5,6$. The bulk states are a sequence of peaks whose amplitude is modulated by the stationary wave in Eq.~\eqref{eq:eigenvaluesgithbingindmodel3}. The two vertical dashed gray lines mark the position of the two boundary potential minima $x=\pm x_B$ [see Eq.~\eqref{eq:grosspitaevsiiequationlattice4}]. The quasi-degenerate boundary states are instead a symmetric and anti-symmetric combination of localized peaks at the two boundary sites. The fact that, in the bulk states, the amplitude of the boundary peaks is nonzero, as well as the fact that the boundary states extend a bit also in the bulk, is ascribed to the fact that, with $x_{\rm min}=6$, the system is not in the strict deep-lattice limit.}
\label{fig:spectrumtightbinding2}
\end{figure*}

In our simulations, we fix $\epsilon=10^{-12}$ and $\gamma=0.1$, and observe that convergence is reached for $100<\overline{n}<400$, for all values of $g$ and $\sigma$ used in this paper. Let us also point out that the specific choice of the \revone{regularized box} kernel \revone{$K(x)$} in Eq.~\eqref{eq:equationinteractionkernel} is motivated by numerical considerations. On one hand, a box potential reduces the numerical complexity of the problem, because it allows for small system lengths $L_x$, and in turn not too large values of $M$, which significantly helps to keep a reasonable numerical complexity of the problem. \revone{Using a kernel from the solution of the heat equation~\cite{PhysRevA.90.043853,weitz2017thermoopticalinteraction,Stein_2022}, or a Gaussian kernel~\cite{PhysRevA.90.043853}, requires indeed larger values of $L_x$, and thus of $M$, especially in the highly nonlocal regime, due to the slower decaying behaviour for large $|x|$}. On the other hand, a sharp box potential (i.e., for $w\rightarrow0$) introduces in the numerical simulations a discontinuity that may spoil numerical convergence. Due to these issues, in order to have a good numerical resolution and a reasonable numerical complexity, we opt to use the regularized box potential.

We mention that a similar numerical method, specifically to solve the two-dimensional Gross-Pitaevskii equation in cylindrical coordinates, was used in Ref.~\cite{PhysRevA.90.043853}.

\section{Spectrum in the deep lattice limit}
\label{appendix:spectrumtightbindinglimit}
In this appendix, we report for the sake of completeness the derivation of the spectrum of Eq.~\eqref{eq:grosspitaevskiiequation2}, in the one-dimensional case, with the lattice potential in Eq.~\eqref{eq:grosspitaevsiiequationlattice4}, and in the linear case ($gN_0=0$). We focus on the first $D_x$ energy levels. As discussed in Sec.~\ref{sec:modelandlinearspectrum}, the spectrum consists of $D_x-2$ states identifying the bulk states, forming the lowest tight-binding energy band, and two states above the band that describe the two boundary states.

In order to find the first $D_x$ states of the spectrum, one can resort to first order degenerate perturbation theory. In the limit $x_{\rm min}/\xi\rightarrow\infty$, the system consists of $D_x$ independent one-dimensional harmonic oscillator wells with potential $V(x)=m\omega^2(x-x_m)^2/2$, where $\{x_m\}$ are the positions of the central point of each potential well of the lattice. By taking each well in its ground state, the spectrum then consists of $D_x$ degenerate levels at energy $\mu_m=\hbar\omega/2$ and eigenfunctions given by the Wannier localized functions $f_m(x)=A_m\,e^{-(x-x_m)^2/2\xi^2}/\sqrt{\xi}\pi^{1/4}$ ($m=1,\ldots,D_x$), where $\{A_m\}$ are proper normalization factors. Out of these $D_x$ states, $L=D_x-2$ states describe Gaussian functions localized at the bulk sites, and the other two are Gaussians localized at the two boundary sites. Of course, for infinitely distant wells, there is no distinction between bulk and boundaries.

Let us now consider $x_{\rm min}/\xi\gg1$ but finite. Here is where bulk and boundary states become clearly distinguished: Reducing the distance between the lattice minima starting from infinity will reduce the potential barriers separating nearest-neighbour wells in the bulk, causing the localized wavefunctions to have an exponentially vanishing but nonzero overlap. At low energy, since the two boundary sites keep the high barrier potentials, the Wannier functions prefer to occupy the bulk sites, while leaving the boundary sites empty.  Let us now focus on the $L=D_x-2$ bulk states. In the basis of the Wannier functions $\{f_m(x)\}$, the effect of reducing $x_{\rm min}/\xi$ can be described by starting from the case of infinitely separated wells, and adding a perturbation $\delta V$ which couples the localized Gaussian $f_m(x)$ only to its nearest-neighbour Gaussians $f_{m\pm1}(x)$. As such, the perturbation $\delta V$ in the basis $f_m(x)\equiv\langle x|f_m\rangle$ has elements given by $\delta V|_{m,m'}\equiv\langle f_m|\delta V|f_{m'}\rangle=-J_0\,\delta_{m',m\pm1}$, where $J_0>0$ quantifies the strength of the nearest-neighbour wavefunctions overlap (i.e., the tight-binding tunneling integral). Now, the perturbation lifts the degeneracy and spectrum becomes $\mu_m\simeq\hbar\omega/2+J_0\lambda_m$, where the corrections $\{\lambda_m\}$ are the eigenvalues of the matrix $\delta V/J_0$.

The eigenvalues $\{\lambda_m\}$ can be found by the following observation. Let us now take $m=1,\ldots,L$, for some $L$, and let us define the matrix $\Lambda_L=-\delta V/J_0+\lambda\mathbb{1}$, where $\mathbb{1}$ is the $L\times L$ identity matrix. Element-wise, $\Lambda_L|_{j,j\pm1}=1$, $\Lambda_L|_{j,j}=\lambda$, and all other elements are zero. The eigenvalues of $-\delta V/J_0$ are the roots of the characteristic polynomial $P_L(\lambda)\coloneqq{\rm det}(\Lambda_L)$. By explicit inspection of the matrix $\Lambda_L$, it can be shown by induction that the following recursion relation holds: ${\rm det}(\Lambda_L)=\lambda\,{\rm det}(\Lambda_{L-1})-{\rm det}(\Lambda_{L-2})$, or in terms of characteristic polynomials
\begin{equation}
P_L(\lambda)=\lambda\,P_{L-1}(\lambda)-P_{L-2}(\lambda) \,\, .
\label{eq:matricesfordeltavminusidenityspectrum4}
\end{equation}
This relation makes sense for $L\geq3$. For $L=2$, one has $P_2(\lambda)=\lambda^2-1$. By introducing the new variable $z=\lambda/2$, i.e., $\lambda=2z$, one has $P_2(z)=4z^2-1$, and the recurrence relation in Eq.~\eqref{eq:matricesfordeltavminusidenityspectrum4} becomes (set $L\rightarrow L+1$)
\begin{equation}
P_{L+1}(z)=2z\,P_L(z)-P_{L-1}(z) \,\, .
\label{eq:matricesfordeltavminusidenityspectrum5}
\end{equation}
The recurrence relation in Eq.~\eqref{eq:matricesfordeltavminusidenityspectrum5}, together with the fact that $P_2(z)=4z^2-1$ is the recurrence relation of the Chebyshev polynomials of the second kind~\cite{zwillinger2011crc}, i.e. $P_L(z)\equiv U_L(z)$. In order to find the eigenvalues, i.e., the roots of $P_L(\lambda)$, one computes the roots of $U_L(z)$ by setting $z=\cos(\theta)$. By further using the expression~\cite{zwillinger2011crc}
\begin{equation}
U_L[\cos(\theta)]=\frac{\sin[(L+1)\theta]}{\sin(\theta)} \,\, ,
\end{equation}
one has $U_L[\cos(\theta)]=0$ for $\theta\equiv\theta_m=\pi m/(L+1)$, but since $\theta={\rm arccos}(z)$, one has $z_m=\cos[\pi m/(L+1)]$. Since by definition $z\in[-1:1]$, one has $\theta\in[0:\pi]$. Since $\lambda=2z$, the roots of $P_L(\lambda)$, ordered such that $\lambda_m\leq\lambda_{m+1}$, can be written as
\begin{equation}
\lambda_m=-2\cos\left(\frac{\pi m}{L+1}\right) \qquad (m=1,\ldots,L) \,\, .
\label{eq:eigenvaluesgithbingindmodel1}
\end{equation}
The eigenvectors $\{\vec v_m\}$ of $-\delta V/J_0$, whose entries $v_{m,j}$ determine the height of the $j$-th wavefunction peak in the $m$-th energy state, are found as customary from the secular equation $(-\delta V/J_0)\vec v_m=\lambda_m\vec v_m$, with $\lambda_m$ as in Eq.~\eqref{eq:eigenvaluesgithbingindmodel1}. Here, we use the following convention on the values of $j$. Due to the presence of the high potential barriers at the lattice boundaries, the effective deep lattice has $L=D_x-2$ sites. In the eigenvalue problem, the values of $j=1,\ldots L$ include only the $D_x-2$ sites of the effective lattice (i.e., excluding the boundaries), where the height of the $j$-th wavefunction peak $v_{m,j}$ can be nonzero. We will then identify the two boundary sites by extending the domain to $j=0$ and $j=L+1$, where by convention $v_{m,j}$ is identically zero on these two sites.

The spatial $j$-dependence of $\{\vec v_m\}$ can be found in the following way. Let us define $v_{m,k}=\sum_je^{ikj}v_{m,j}$, then one has from the element-wise secular equation, which is $-v_{m,j-1}-v_{m,j+1}-\lambda_mv_{m,j}=0$, the expression $\sum_k[2\cos(k)+\lambda_m]e^{-ikj}v_{m,k}=0$. Because of the presence of the oscillating terms $e^{-ikj}$, this relation is satisfied if $[2\cos(k)+\lambda_m]v_{m,k}=0$, for each $k$. This means that one has $v_{m,k}\neq0$ when $2\cos(k)=-\lambda_m$, i.e., by using Eq.~\eqref{eq:eigenvaluesgithbingindmodel1}, when $k=\pm k_0=\pm\pi m/(L+1)$. Otherwise, when $2\cos(k)\neq-\lambda_m$, one has $v_{m,k}=0$. Going back to real space $j$, the $j$-dependence of $\vec v_m$ is
\begin{equation}
v_{m,j}=v_{m,k_0}e^{i\pi mj/(L+1)}+v_{m,-k_0}e^{-i\pi mj/(L+1)} \,\, ,
\label{eq:eigenvaluesgithbingindmodel2}
\end{equation}
for $j,m=1,\ldots,L$. As said before, we require that $v_{m,j}$ is identically zero on the extended domain for $j=0$ and $j=L+1$. In Eq.~\eqref{eq:eigenvaluesgithbingindmodel2}, this boundary condition yields $v_{m,k_0}=-v_{m,-k_0}$, which can be taken as $v_{m,k_0}=-iv_0/2$ to have a real $v_{m,j}$, and then we can write Eq.~\eqref{eq:eigenvaluesgithbingindmodel2} as a stationary wave
\begin{equation}
v_{m,j}=v_0\,\sin\left(\frac{\pi mj}{L+1}\right) \,\, .
\label{eq:eigenvaluesgithbingindmodel2bis0}
\end{equation}
The constant $v_0$ in Eq.~\eqref{eq:eigenvaluesgithbingindmodel2bis0} is chosen to ensure the unit normalization of the vector $\vec v_m$. In the case of a symmetric potential $V(x)=V(-x)$, as in the case of Eq.~\eqref{eq:grosspitaevsiiequationlattice4}, Eq.~\eqref{eq:eigenvaluesgithbingindmodel2bis0} is modified by shifting $j\rightarrow j-(L+1)/2$, i.e.
\begin{equation}
v_{m,j}=\left\{\begin{array}{ll}
\displaystyle{v_0\cos\left(\frac{\pi mx_j}{L+1}\right)} &\quad \mbox{($m$ odd)}\\\\
\displaystyle{v_0\sin\left(\frac{\pi mx_j}{L+1}\right)} &\quad \mbox{($m$ even)}
\end{array}\right. \,\, ,
\label{eq:eigenvaluesgithbingindmodel3}
\end{equation}
where $x_j=j-(L+1)/2$ with $j=1,\ldots,L$ is the position along the $x$-axis of the $j$-th potential minimum (in units of $x_{\rm min}$).

The first $D_x$ eigenvalues $\{\mu_m\}$ and real wavefunctions $\{\psi_m(x)\}$ for the case of the lattice potential discussed in Sec.~\ref{sec:lattices} ($D_x=6$ and $x_{\rm min}=6$) are shown in Figs.~\ref{fig:spectrumtightbinding1} and~\ref{fig:spectrumtightbinding2}, respectively. The spectrum consists of $D_x-2$ bulk energy levels as $\mu_m\simeq\mu_0+J_0\lambda_m$, where $\mu_0$ and $J_0$ quantify the zero-point and bandwidth, and $\lambda_m$ is as in Eq.~\eqref{eq:eigenvaluesgithbingindmodel1}, with $L=D_x-2$. Small deviations from the predicted behaviour are due to the fact that, with $x_{\rm min}=6$, the system is not in the deep lattice limit. Also, the fact that $\mu_0\neq\hbar\omega/2$ is a consequence of the fact that the lattice wells are not perfectly harmonic (they are approximately only at their center). Above the tight-binding band of bulk states, the two quasi-degenerate boundary states (red points) are found. These two states form a ``boundary double well'', where the lowest boundary state is a symmetric combination of two localized functions at the boundary sites, whereas the highest boundary state is instead an anti-symmetric combination (see also Fig.~\ref{fig:spectrumtightbinding2}).

For completeness, we mention that, for $m>D_x$, the higher bands are encountered. The wavefunctions in these bands are characterized by an increasing number of nodes appearing at the minima of the potential wells, i.e., some of the Wannier localized functions are not Gaussians but rather higher Hermite-Gauss modes.


%

\end{document}